\documentclass{aa}  

\usepackage{graphicx}
\usepackage{lipsum}
\usepackage[colorlinks]{hyperref}
\usepackage{color, colortbl} 
\usepackage{longtable} 
\usepackage{lscape}
\usepackage{txfonts}

\newcommand{\kms}{\,km\,s$^{-1}$}
\newcommand{\msun}{\,M$_\odot$}

\begin{document} 

\makeatletter
\setlength{\aa@afterheadboxskip}{0.5cm}    
\makeatother

\title{ALMA Lensing Cluster Survey: Dust mass measurements as a function of redshift, stellar-mass and star formation rate, from $z=1$ to $z=5$}

 \author{Jean-Baptiste Jolly\inst{1,2} \and Kirsten Knudsen\inst{2} \and Nicolas Laporte\inst{3} \and Andrea Guerrero\inst{4} \and Seiji Fujimoto\inst{5,6} \and Kotaro Kohno\inst{7,8} \and Vasily Kokorev\inst{9} \and Claudia del P. Lagos\inst{10} \and Thiébaut-Antoine Schirmer\inst{2} \and  Franz Bauer\inst{11,12,13} \and Miroslava Dessauge-Zavadsky\inst{14} \and Daniel Espada\inst{15,16} \and Bunyo Hatsukade\inst{7} \and Anton M. Koekemoer\inst{17} \and Johan Richard\inst{18}  \and Fengwu Sun\inst{19} \and John F. Wu\inst{20,21} }

   \institute{Max-Planck-Institut für extraterrestrische Physik, 85748 Garching, Germany   
    \and Department of Space, Earth and Environment, Chalmers University of Technology, SE-412 96 Gothenburg, Sweden  \\   
    \email{jbjolly@mpe.mpg.de}
         \and
        Aix Marseille Université, CNRS, CNES, LAM (Laboratoire d’Astrophysique de Marseille), UMR 7326, 13388 Marseille, France
        \and
        Astronomy Department, Universidad de Concepci\'{o}n, Barrio Universitario S/N, Concepci\'{o}n, Chile
        \and
        Cosmic Dawn Center (DAWN), Denmark 
        \and
        Niels Bohr Institute, University of Copenhagen, Jagtvej 128 DK-2200 Copenhagen N, Denmark
        \and
        Institute of Astronomy, Graduate School of Science, The University of Tokyo, 2-21-1 Osawa, Mitaka, Tokyo 181-0015, Japan
        \and
        Research Center for the Early Universe, Graduate School of Science, The University of Tokyo, 7-3-1 Hongo, Bunkyo-ku, Tokyo 113-0033, Japan
        \and Kapteyn Astronomical Institute, University of Groningen, PO Box 800, 9700 AV Groningen, The Netherlands
        \and
        International Centre for Radio Astronomy Research (ICRAR), M468, University of Western Australia, 35 Stirling Hwy, Crawley, WA 6009, Australia
        \and
        Instituto de Astrof{\'{\i}}sica, Facultad de F{\'{i}}sica, Pontificia Universidad Cat{\'{o}}lica de Chile, Campus San Joaquín, Av. Vicuña Mackenna 4860, Macul Santiago, Chile, 7820436
        \and 
        Centro de Astroingenier{\'{\i}}a, Facultad de F{\'{i}}sica, Pontificia Universidad Cat{\'{o}}lica de Chile, Campus San Joaquín, Av. Vicuña Mackenna 4860, Macul Santiago, Chile, 7820436 
        \and 
        Millennium Institute of Astrophysics, Nuncio Monse{\~{n}}or S{\'{o}}tero Sanz 100, Of 104, Providencia, Santiago, Chile
        \and
        Observatoire de Genève, Université de Genève, 51 Ch. des Maillettes, 1290 Versoix, Switzerland
        \and
        Departamento de F\'{i}sica Te\'{o}rica y del Cosmos, Campus de Fuentenueva, Edificio Mecenas, Universidad de Granada, E-18071, Granada, Spain
        \and
        Instituto Carlos I de F\'{i}sica Te\'{o}rica y Computacional, Facultad de Ciencias, E-18071, Granada, Spain
        \and
        Space Telescope Science Institute, 3700 San Martin Dr., Baltimore, MD 21218, USA
         \and
         CRAL, Observatoire de Lyon, Universit\'e Lyon 1, 9 Avenue Ch. Andr\'e, F-69561 Saint Genis Laval Cedex, France
         \and
         Steward Observatory, University of Arizona, 933 N. Cherry Avenue, Tucson, 85721, USA
         \and
         Space Telescope Science Institute, 3700 San Martin Dr., Baltimore, MD 21218, USA
        \and
        Department of Physics \& Astronomy, Johns Hopkins University, 3400 N Charles St, Baltimore, MD 21218
         }

   \date{Accepted November 15, 2024 (Submitted to A\&A)}

  \abstract
   {Understanding the dust content of galaxies, its evolution with redshift and its relationship to stars and star formation is fundamental for our understanding of galaxy evolution. Dust acts as both a catalyst of star formation and as a shield for star light. State-of-the-art millimeter ground-based facilities like ALMA have made dust observation ever more accessible, even at high redshift. However dust emission is typically very faint, making the use of stacking techniques instrumental in the study of dust in statistically sound samples.}
       {Using the ALMA Lensing Cluster Survey (ALCS) wide-area band-6 continuum dataset ($\sim\,$110 arcmin$^2$ across 33 lensing clusters), we aimed at constraining the dust mass evolution with redshift, stellar mass and star formation rate (SFR).}
       {After binning sources according to redshift, SFR and stellar mass --extracted from an HST-IRAC catalog-- we performed a set of continuum stacking analyses in the image domain using \textsc{LineStacker} on sources between $z=1$ and $z=5$, further improving the depth of our data. The large field of view provided by the ALCS allows us to reach a final sample of $\sim4000$ galaxies with known coordinates and SED-derived physical parameters. We stack sources with SFR between $10^{-3}$ and $10^{3}$ M$_\odot$ per year, and stellar mass between $10^{8}$ and $10^{12}$ M$_\odot$, splitting them in different stellar mass and SFR bins. Through stacking we retrieve the continuum 1.2\,mm flux, a known dust mass tracer, allowing us to derive the dust mass evolution with redshift and its relation with SFR and stellar mass.}
       {We observe clear continuum detections in the majority of the subsamples. From the non detections we derive 3-$\sigma$ upper limits. We observe a steady decline in the average dust mass with redshift. Moreover, sources with higher stellar mass or SFR have higher dust mass on average, allowing us to derive scaling relations. Our results are mostly in good agreement with models at $z\sim1$ -- $3$, but indicate typically lower dust-mass than predicted at higher redshift. }
        {}

\titlerunning{ALCS: dust mass measurements}
\authorrunning{Jean-Baptiste Jolly et al.}

\maketitle

\newpage
%
\section{Introduction}

Dust impacts both the direct evolution of galaxies as well as their observations in multiple ways. It is thought to be the main catalyst of H$_2$ formation \citep{Wakelam2017}, one of the main components of the molecular clouds in the interstellar medium (ISM) and hence one of the principal drivers of star formation \citep[e.g.][]{Scoville2013}. In addition to the direct effect on the physical properties of galaxies, dust also plays a critical role in astrophysical observations. The dust grains attenuating the starlight are consequently heated, and in turn re-radiate the energy at longer wavelengths, this is the so-called dust-obscured star formation \citep[see for e.g.][]{Casey2014,Hodge2020,Zavala2021}. As a consequence, the light coming from star-forming regions is typically dominated by dust continuum emission. The relationship between dust mass and other physical characteristics of galaxies have been intensely studied, for example, by studying dust-to-gas (atomic and molecular) or dust-to-metal ratios \citep[e.g.][]{Hunt2005,Draine2007,Engelbracht2008,Galametz2011,Magrini2011,Saintonge2013,Remy-Ruyer2014,Combes2018,Li2019,Shapley2020,Tacconi2020,Birkin2021,Popping2022,Popping2023}, or dust mass versus star formation rate (SFR) \citep[e.g.][]{DaCunha2010,Casey2012,Santini2014,Dudzeviciute2020,Dudzeviciute2021}.

Recent studies have shown that the proportion of dust-obscured star formation has evolved with redshift \citep{Bouwens2020}. It has been estimated to be a dominant fraction during the peak of the cosmic star formation history. However, it appears to have accounted for only $\sim$ 20\% -- 25\% of the total star formation at redshift $z=6$ -- $7$ \citep[see][]{Zavala2021}. Similarly, the overall cosmic dust density has been evidenced to peak around $z\sim 1-2$ with a rapid decline at higher redshift \citep[see for example][]{Driver2018,Magnelli2020,Pozzi2020,Dudzeviciute2021}. Furthermore, massive dust reservoirs have been detected in massive high-redshift galaxies \citep[e.g.][]{Bertoldi2003,Valiante2009,Venemans2012,Watson2015,Laporte2017,Tamura2019,Tamura2023}. While this is probably not the case for more typical, less massive high-redshift galaxies, it highlights the importance of studying the evolution of dust mass with redshift.

Millimeter or sub-millimeter emission has been proposed to be a tracer of dust mass \citep{Scoville2014,Scoville2016,Scoville2017}, with the assumption that the emission is optically thin and measured far from the peak of the dust spectral energy distribution. The Atacama Large Millimeter/Submillimeter Array (ALMA) has become instrumental in the quest to study dust mass in lower-mass galaxies and its evolution in the mm and sub-mm bands. With large surveys and high-redshift single target observations becoming progressively more accessible \citep[e.g.][]{Knudsen2016,Walter2016,Gonzalez-Lopez2017,Gonzalez-Lopez2020,Laporte2017,Bethermin2020,Aravena2020} the evolution of dust is studied in increasingly statistically sound samples.  

However, observations of high-redshift galaxies are typically biased by construction towards the observation of the brightest sources (Malmquist bias). To draw a complete picture of dust evolution with redshift it is necessary to also study galaxies with lower intrinsic luminosities. Gravitational lensing can be used as a tool to enhance the signal from faint galaxies without the need for excessive integration time. Similarly, tools like stacking can statistically improve the signal-to-noise ratio (S/N) drastically when studying large samples. With the help of both gravitational lensing and stacking, it is hence possible to push the limit of observations toward sources with lower intrinsic dust luminosity.

In this paper we present the stacking analysis of 10386 gravitationally lensed galaxies from the ALMA Lensing Cluster Survey (ALCS) at $z>1$ (cluster and field sources at $z\lesssim1$ are studied in a separate paper: \citet{Guerrero}. By binning galaxies by redshift, and further splitting them according to their stellar mass or SFR, we study the evolution of the dust mass with redshift, and its scaling relation with stellar mass and SFR. Besides, we also integrate the contribution from all galaxies --in each redshift bin-- to assess the evolution of the cosmic dust density.

In Section \ref{sec:data} we describe the overall data set, the catalog and the different subsamples. In Section \ref{sec:method} we describe both the stacking method, performed using \textsc{LineStacker} \citep{Jolly2020}, as well as the processes involved in the dust mass calculation. In Section \ref{sec:results} we present our results and discuss them in Section \ref{sec:discussion}. Finally in Section \ref{sec:summary} we present a summary of this paper. In the appendix we show alternative stacking procedures as well as some extra details on the main analysis.

Throughout the paper, we assume a $\Lambda$CDM cosmology with $\Omega_m = 0.3$, $\Omega_\lambda = 0.7$ and $H_0=70$ \kms Mpc$^{-3}$. All magnitudes are quoted in the AB system, such that M$_{\mathrm{AB}}$ = 23.9-2.5 log$_{10}$ (S$_\nu$ [$\mu$Jy]).

\section{Data and sample} \label{sec:data}

\subsection{ALCS} \label{sec:ALCS}

The ALMA Lensing Cluster Survey (ALCS) is a large ALMA program accepted in cycle 6 (Project ID: 2018.1.00035.L; PI: K. Kohno). It observed 33 lensing clusters in band 6 ($\lambda\sim1.2\,$mm), spanning $\sim$\,110\,arcmin$^2$ (primary beam (PB) $>0.5$). The clusters are distributed as follow: 16 from RELICS \citep[the Reionization Lensing Cluster Survey][]{Coe2019}, 12 from CLASH \citep[the Cluster Lensing And Supernova survey with Hubble][]{Postman2012}, and 5 from the Frontier Fields survey \citep{Lotz2017}. Observations were carried out between December 2018 and December 2019 (cycles 6 and 7) in compact array configurations C43-1 and C43-2, in a double frequency window setup, observing both from 250.0 to 257.5\,GHz and from 265.0 to 272.5\,GHz, for a total bandwidth of 15\,GHz. When available, the ALCS data were concatenated with existing ALMA data, notably the ALMA Frontier Fields Survey (Project ID: 2013.1.00999.S, PI: Bauer and Project ID; 2015.1.01425.S: PI: Bauer). The data were reduced and calibrated using the Common Astronomy Software Applications \citep[\textsc{CASA,}][]{CASA} package version 5.4.0 for the 26 clusters observed in cycle 6 and v5.6.1 for the remaining clusters observed in cycle 7. Throughout this paper, we use natural-weighted, primary-beam-corrected, and {\it uv}-tapered continuum maps, with a tapering parameter of 2\,arcsec (the full width at half maximum of the synthesized beam is 2\,arcsec; with a corresponding pixel size of 0.16 arcsec). {\it uv}-tapered maps have been chosen over natural resolution maps to ensure beam-size homogeneity of the different images when stacking. The average RMS of the maps is $\sim63\, \mathrm{\mu}$Jy/beam, see Table \ref{tab:cluster_sources} for the detailed RMS of each map. Full description of the survey can be found in \citet{KohnoALCS}.

\subsection{Source catalogue} \label{sec:cat}

We extracted the positions, redshifts, physical characteristics (SFRs and stellar masses) as well as lensing magnifications of galaxies at $1\leq z \leq 5$ from the HST-IRAC catalog presented in \citet{Kokorev2022}. When available, spectroscopic redshifts are used in place of photometric redshifts. While the SFRs and stellar masses extracted from the \citet{Kokorev2022} catalog are not corrected for magnification, the SFRs and stellar masses presented in this paper are always the intrinsic ones, i.e. corrected for magnification.

From the full catalog we select only the sources that (the number of remaining sources is indicated in italic after each step): (i) have defined redshifts and magnifications, and are not tagged with bad photometry (bad\_phot $\neq 1$) -- \textit{1888977} (ii) have redshift uncertainties ($|z_{160}-z_{840}|$) below 0.4 -- \textit{76728} (iii) have $10^8 \leq \mathrm{M}_* / \mathrm{M}_\odot \leq 10^{12}$ and $0.001 \leq \mathrm{SFR} / ( \mathrm{M}_\odot  / \mathrm{year}) \leq 10^3$ -- \textit{51355} (iv) have H-band magnitudes above 24, to avoid contamination from blue faint galaxies -- \textit{50983} (v) have associated PB values higher than 0.5 -- \textit{13403} (vi) have a $z<5$ -- \textit{12980} (vii) have a magnification factor below 100 -- \textit{12967} (viii) have a $z\geq 1$ -- \textit{4103}, as lower redshift sources as well as cluster sources were presented in a separate paper, \cite{Guerrero}. In total, our full sample finally adds up to a total of 4103 sources. Table \ref{tab:cluster_sources} presents the distribution of sources in the 33 clusters observed in the ALCS as well as the RMS of each map. 

\begin{figure*}
   \centering
    \includegraphics[width=0.9\columnwidth]{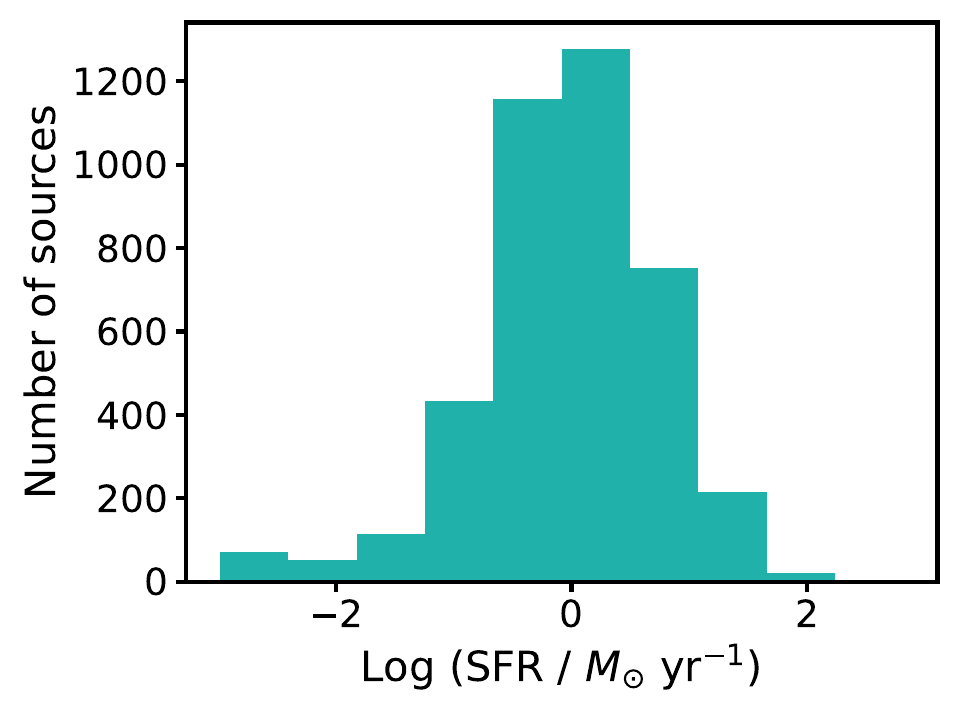}   
    \includegraphics[width=0.9\columnwidth]{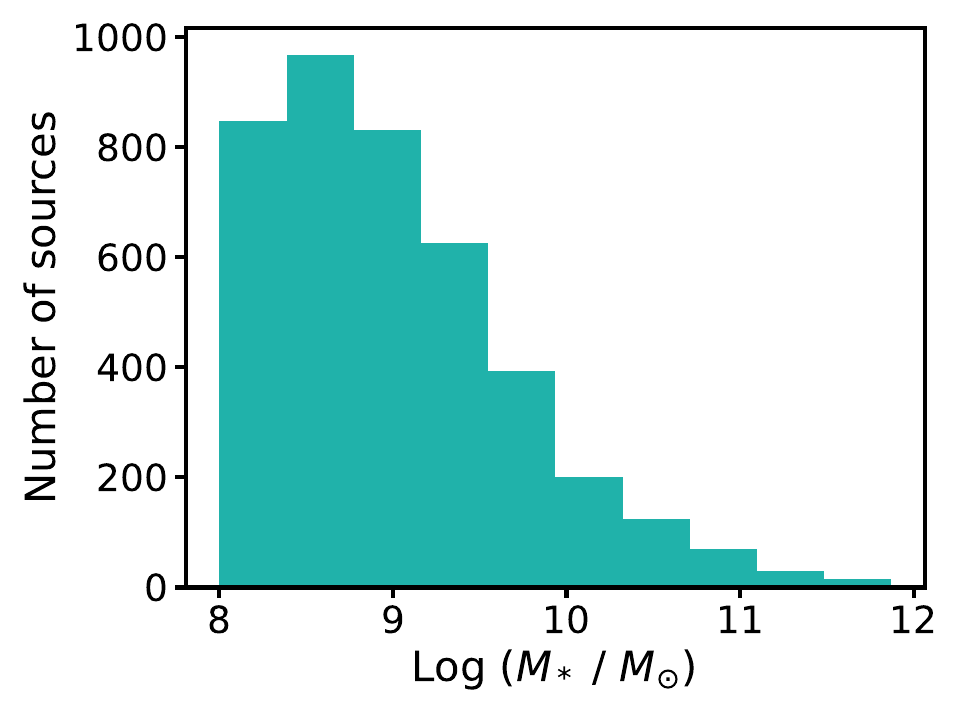}
    \includegraphics[width=0.9\columnwidth]{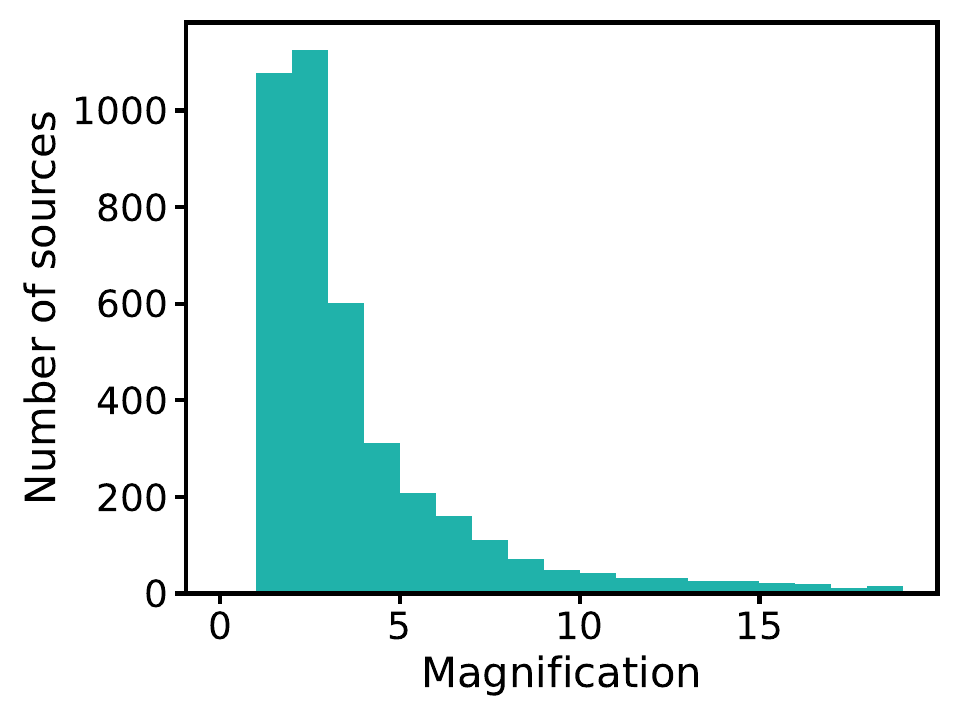}
    \includegraphics[width=0.9\columnwidth]{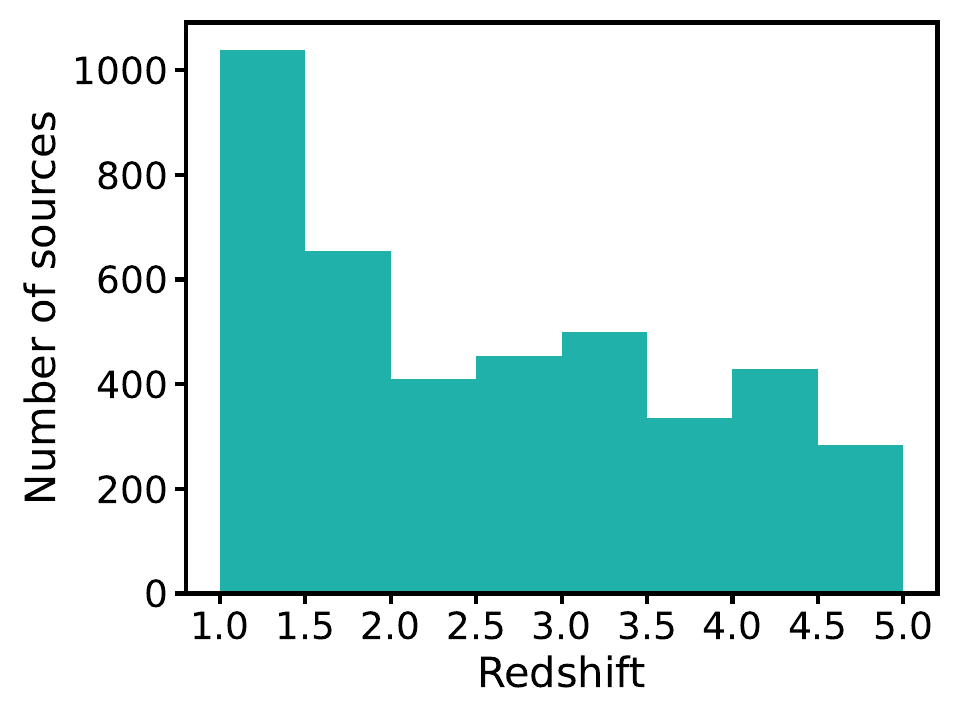}       
      \caption{Distribution of the main physical properties of interest in the whole sample. SFRs and stellar masses are corrected for magnification.
              }
         \label{fig:distrib}
   \end{figure*}

\begin{figure}
   \centering
    \includegraphics[width=0.9\columnwidth]{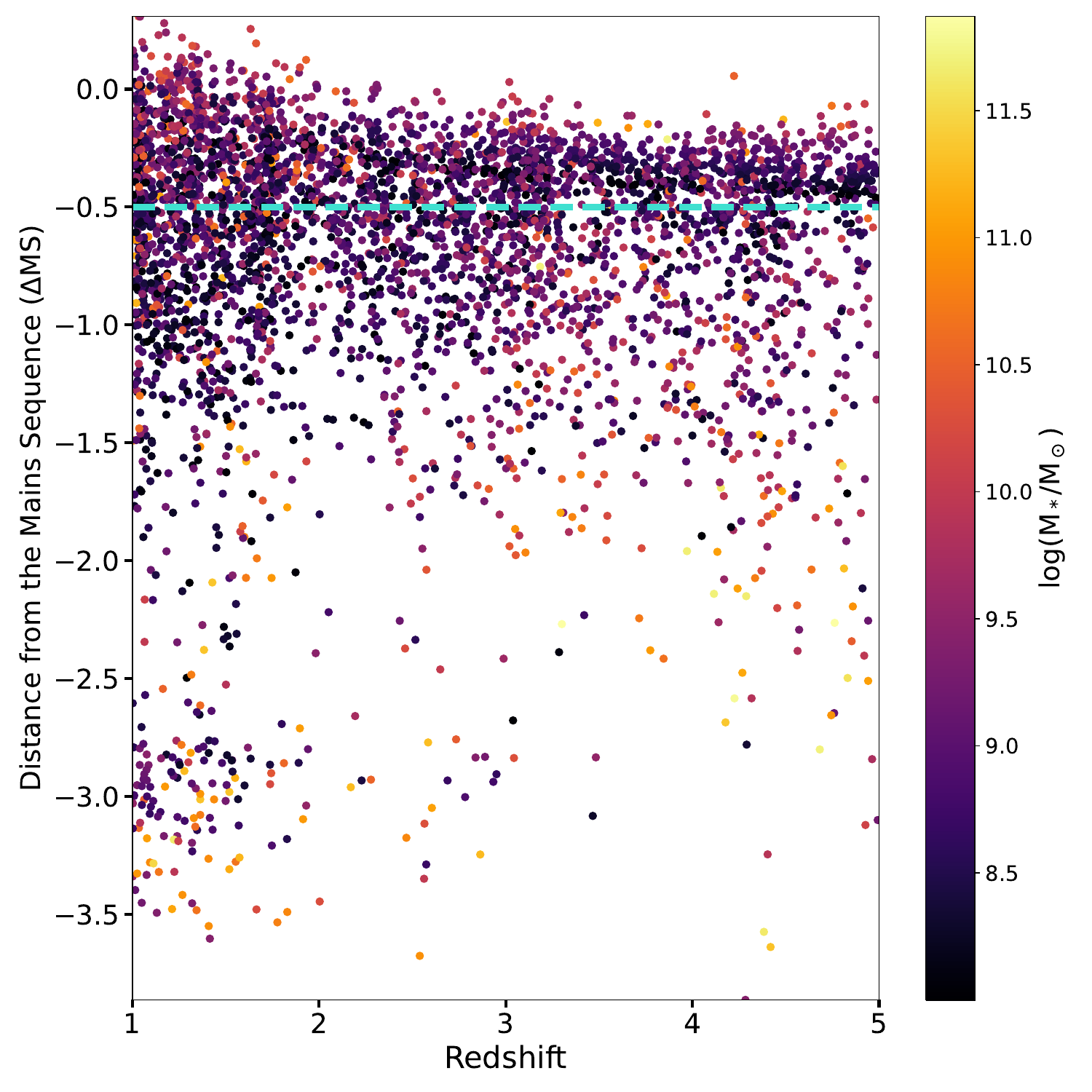}
      \caption{Distance from the main sequence $\Delta({\mathrm{MS}}) = \mathrm{log}(\mathrm{SFR}_{\mathrm{MS}}) - \mathrm{log}(\mathrm{SFR})$ \citep[from][]{Speagle2014} as a function of redshift, for the galaxies in the full sample. Points are color coded for their stellar mass. A dashed line at $\Delta({\mathrm{MS}})=-0.5$ highlights the region of exclusion of quiescent galaxies (see Appendix).}
         \label{fig:dMS}
   \end{figure}

Figure \ref{fig:distrib} shows the distributions of stellar mass, SFR, magnification and redshift in the final sample. Figure \ref{fig:dMS} shows the distance of each galaxy from the galaxy main sequence (MS) as a function of redshift -- $\Delta({\mathrm{MS}}) = \mathrm{log}(\mathrm{SFR}_{\mathrm{MS}}) - \mathrm{log}(\mathrm{SFR})$ with $\mathrm{SFR}_{\mathrm{MS}}$ the expected SFR for a galaxy with the same stellar mass and on the MS \citep[according to][]{Speagle2014}, and SFR the actual SFR of the galaxy. The sample shows an over-density of quiescent galaxies \citep[as also shown in ][]{Guerrero}. To assess the impact of quiescent galaxies on the stack we also stacked every sample after first excluding galaxies with $\Delta$MS\,$<-0.5$, see Appendix.

\begin{figure}
   \centering
    \includegraphics[width=0.9\columnwidth]{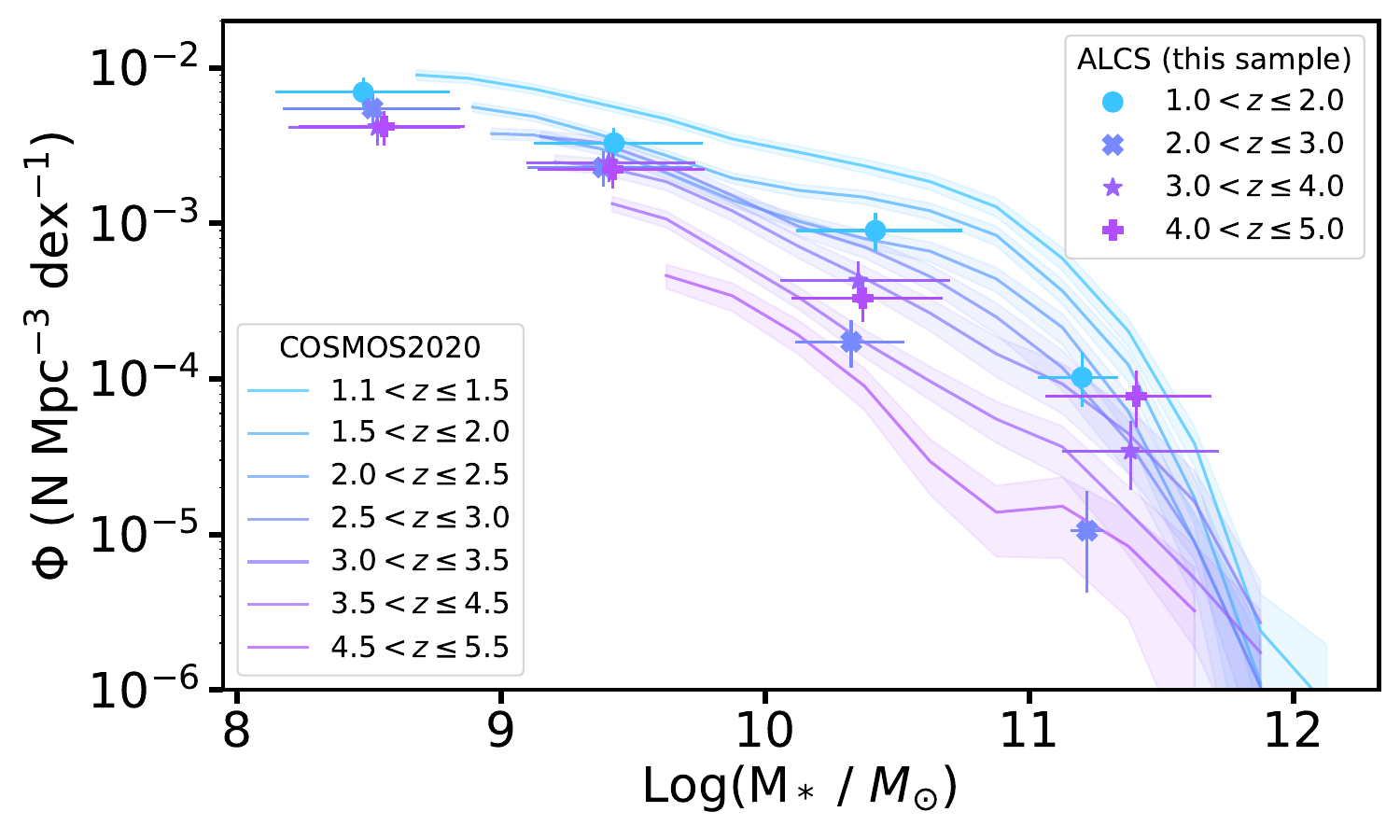}      
      \caption{Comparison between the SMF of the COSMOS2020 sample \citep{Weaver2022,Weaver2023} and the SMF of the sources studied in this paper. The SMF is obtained by counting the galaxies in each redshift bin and dividing by the corresponding volume, corrected for the mean magnification of the sources in each bin, for each cluster separately. The error bars correspond to the 16$^{\mathrm{th}}$ to 84$^{\mathrm{th}}$ percentile of the stellar mass of the sources in the sample, and the impact of the $20\%$ error on the magnification on the total volume.} 
         \label{fig:SMF}
   \end{figure}

To further evaluate the completeness and reliability of the sample we compare its stellar mass function (SMF) to the one derived in the COSMOS2020 sample \citep{Weaver2022,Weaver2023}. To derive the SMF of our sample we count --for each cluster separately-- the number of galaxies in a given redshift bin and divide by the corresponding volume, corrected for the mean magnification of the sources in the studied bin. The comparison is shown on Figure \ref{fig:SMF}. The studied sample seems to present an over-density of sources at $z>4$, and an under-estimation of sources in the $2<z \leq3$ bin, especially at high-masses. This might be due to the miss-identification of low redshift sources as high-redshift ones, as also discussed in \citet{Guerrero}. \citet{Kokorev2022} compared the spectroscopic redshifts, known for $\sim7000$ galaxies, to the photometric redshifts they derived: while $\sim80\%$ of the redshifts agree reasonably well, the remaining $\sim20\%$ present large to sometimes catastrophic ($\Delta z >3$) errors, mainly due to the confusion of the Lyman, Balmer, and 4000\,$\AA$ breaks. In addition, an under-evaluation of the survey volume (coming from an over-evaluation of the corresponding magnification) might also artificially boost the source density. The potential impact of the derived SMF are further discussed in Section \ref{sec:discussion}.

\begin{table}
\caption{\label{tab:cluster_sources} Number of stacked sources in each cluster, and maps RMS.}
\centering   
 \begin{tabular}{lcc}
 \hline \hline
Cluster name & Number of stacked sources & Maps RMS \\ 
 & & ($\mu$Jy/beam)\\
\hline
HFF clusters: \\
Abell2744 & 418 & 51 \\
Abell370 & 346 & 47 \\
AbellS1063 & 369 & 53 \\
MACSJ0416.1-2403 & 473 & 55 \\
MACSJ1149.5+2223 & 325 & 64 \\
[0.3cm]

CLASH clusters:\\
Abell209 & 74 & 63\\
Abell383 & 63 & 61\\
MACS0329.7-0211 & 62 & 71 \\
MACS0429.6-0253 & 37 & 92 \\
MACS1115.9+0129 & 51 & 63 \\
MACS1206.2-0847 & 127 & 53 \\
MACS1311.0-0310 & 38 & 62 \\
MACS1423.8+2404 & 101 & 65 \\
MACS1931.8-2635 & 80 & 56 \\
MACS2129.4-0741 & 116 & 47 \\
RXJ1347-1145 & 95 & 53\\
RXJ2129.7+0005 & 31 & 40 \\
[0.3cm]

RELICS clusters:\\
Abell2163 & 11 & 50 \\
Abell2537 & 52 & 69 \\
Abell3192 & 91 & 73 \\
AbellS295 & 39 & 74\\
ACTCLJ0102-49151 & 253 & 72\\
MACSJ0035.4-2015 & 72 & 52 \\
MACSJ0159.8-0849 & 69 & 63\\
MACSJ0257.1-2325 & 67 & 83 \\
MACSJ0417.5-1154 & 135 & 84 \\
MACSJ0553.4-3342 & 157 & 65 \\
PLCKG171.9-40.7 & 28 & 73 \\
RXCJ0032.1+1808 & 81 & 71\\
RXCJ0600.1-2007 & 66 & 57 \\
RXCJ0949.8+1707 & 38 & 62 \\
RXCJ2211.7-0350 & 52 & 78 \\
SMACSJ0723.3-7327 & 86 & 66 \\
[0.3cm]
 & Total: 4103 & Average: 63 \\
\hline
\end{tabular}

\end{table}

\subsection{Subsamples} \label{sec:sub-samples}

To study the evolution of dust mass across redshift we separated the sources in four redshift bins: $1\leqslant z<2$, $2\leqslant z<3$, $3\leqslant z<4$ and $4\leqslant z<5$. We additionally split sources according to either their SFRs or stellar masses in the following bins: (i) SFR, (upper and lower limits in \msun\,yr$^{-1}$):  $0.001\leqslant$ SFR $\leqslant1$, $1<$ SFR $\leqslant10$, $10<$ SFR $\leqslant50$, $50<$ SFR $\leqslant100$ and $100<$SFR $\leqslant1000$; (ii) stellar mass, (M$_*$, upper and lower limits in \msun):  $10^8\leqslant M_* \leqslant 10^9$, $10^9< M_* \leqslant\,10^{10}$, $10^{10}< M_* \leqslant 10^{11}$ and $10^{11}< M_* \leqslant 10^{12}$. The total number of sources in each subsample can be found on Table \ref{tab:subsample_sources}. Further, the distribution of SFR and stellar mass in each subsample can be found in the appendix, on Figures \ref{fig:mass_subsample_hist} and \ref{fig:sfr_subsample_hist}.

One should note that sources individually detected in the ALCS data are not excluded from the samples. We decided to discriminate solely on the properties mentioned above to derive average properties of the population, and, in that regard, the 1.2\,mm continuum flux density of individual sources should not be a criterion of exclusion from our samples. To assess the impact of the 1.2\,mm flux distribution in our samples, we perform median stacking as well as bootstrapping analyses, alongside the main mean-stacking analyses, see Sections \ref{sec:errors} and \ref{sec:median_stacking}.

\begin{table*}
\caption{\label{tab:subsample_sources} Number of stacked sources in each subsample.}
\centering   
 \begin{tabular}{lccccc}
 \hline \hline
& $1\leqslant z<2$ & $2\leqslant z<3$ & $3\leqslant z<4$ & $4\leqslant z<5$ & Total\\ 
\hline

8$\leqslant$log($M_*$)$\leqslant$9 & 1004 & 521 & 432 & 375 & 2332 \\
9<log($M_*$)$\leqslant$10 & 508 & 295 & 309 & 256 & 1368 \\ 
10<log($M_*$)$\leqslant$11 & 154 & 43 & 84 & 60 & 341 \\ 
11<log($M_*$)$\leqslant$12 & 27 & 4 & 11 & 20 & 62 \\ 
\hline
0.001$\leqslant$SFR$\leqslant$1 & 1046 & 441 & 306 & 250 & 2043 \\ 
1<SFR$\leqslant$10 & 529 & 377 & 455 & 394 & 1755 \\ 
10<SFR$\leqslant$50 & 113 & 43 & 67 & 58 & 281 \\ 
50<SFR$\leqslant$100 & 5 & 1 & 1 & 5 & 12 \\ 
100<SFR$\leqslant$1000 & - & 1 & 7 & 4 &  12 \\ 

\hline
\end{tabular}
\end{table*}

\section{Methods} \label{sec:method}

\subsection{Stacking} \label{sec:stacking}

Stacking was done using \textsc{LineStacker} \citep{Jolly2020}, in single channel mode (to perform continuum stacking), on {\it uv}-tapered images (see Section \ref{sec:ALCS}). Sources were stacked pixel to pixel using mean stacks without weights (median stacks are also performed, see Section \ref{sec:median_stacking}). Stacking was operated with stamp sizes of 9.76 $\times$ 9.76 arcsec$^2$ (61 $\times$ 61 pixels). Sources were spatially aligned using the position extracted from the \citet{Kokorev2022} catalog. Stacked flux in Jy was retrieved by integrating the continuum flux in a central circular region of the stack stamp, with a radius $R_{\rm integ}=2''$ (12.5 pixels), corresponding to the synthesized beam size, and dividing it by the beam size in pixel units. 

To compute the RMS associated to each stacked cube a random set of source-free coordinates are drawn for each stacking position. An empty stack is then generated from the set of random coordinates \citep[see][]{Jolly2020} with the same characteristics as the normal stacks (i.e. stamp size and number of targets). This process is performed 1000 times for each subsample. The standard deviation across all stack stamps is then computed and used to derive the RMS associated to each stack cube.

When the flux in the central region of the stack stamp is lower than 3 times its associated RMS, a $3 \sigma$ upper limit is used in place of the integrated flux to compute $M_{\mathrm{dust}}$.

\subsection{Magnification corrections} \label{sec:magnif}

The averaged 1.2\,mm fluxes extracted from each stack maps are corrected for the mean magnification of sources in the stack subsample, the final flux is hence computed as: $F_{final} = \frac{F_{stack} }{\overline{\mu}}$, where $F_{final}$ is the final flux used to compute the dust mass (see Section \ref{sec:dust}), $F_{stack}$ is the flux obtained from the stacked map, and $\overline{\mu}$ is the average magnification of each source in the subsample. Alternatively, each source could be corrected for its magnification before stacking: $F_{final} = F_{\mathrm{corrected\ stack}}$, where $F_{\mathrm{corrected\ stack}}$ is the flux obtained from the stack of the individually magnification-corrected maps: $\mathrm{Map}_{\mathrm{stack}} = \sum_{i=1}^n \frac{\mathrm{Stamp}_{i}}{\mu_{i}}$, where $\mathrm{Map}_{\mathrm{stack}}$ is the stacked map obtained from individually magnification-corrected stamps (from which $F_{\mathrm{corrected\ stack}}$ is obtained), $n$ is the number of sources in the subsample, $\mathrm{Stamp}_{i}$ is the map-stamp associated to source $i$ and $\mu_i$ is the magnification associated to source $i$. The reason why we decided to correct for the average magnification of the sample after stacking is the following: the signal in each stack stamp consists of both the flux from the source, which is gravitationally magnified, and some noise, which is independent of the magnification. By correcting each stack stamp for its associated  magnification, one corrects both the flux from the source and the foreground noise. The efficiency of stacking is based on the fact that the noise is random and approximately similar in every stack stamp. By down-scaling the noise in each stamp by a different number (unrelated to the noise properties), one might effectively reduce the efficiency and reliability of the stacking analysis. However, by correcting the final stack with the average magnification, the analysis would be biased toward high-magnification sources. Correcting each stamp individually pre-stacking should avoid this bias, we hence repeated our analysis with this correction method. The 1.2\,mm fluxes derived in both cases are very consistent, with values overlapping within the error margins.

\subsection{Measuring $M_{\rm dust}$} \label{sec:dust}

To derive dust masses using the 1.2\,mm continuum flux extracted from each stacking stamp, we use a single modified blackbody curve, under the approximation of an optically thin regime, following \citet{Kovacs2010} and \citet{Magnelli2020} we define $M_{\rm dust}$ as:

\begin{equation} \label{eq:dust}
    M_{\rm dust}=\frac{5.03 \times 10^{-31} ( S_{\nu_{\rm obs}} / f_{\mathrm{CMB}}) D_{\rm L}^2 }{(1+z)^4 B_{\nu_{\rm obs}}(T_{\rm obs})  \kappa_{\nu_0}}  \left(\frac{\nu_0}{\nu_{\rm rest}}\right)^{\beta},
\end{equation}

\noindent where $M_{\rm dust}$ is the dust mass in \msun, $S_{\nu_{\rm obs}}$ is the flux at the observed frequency in Jy (and corrected for the average magnification, as explained in Section \ref{sec:magnif}), $\nu_{\rm obs}=\nu_{\rm rest}/(1+z)$ is the observed frequency, $f_{\mathrm{CMB}}$ is the correction factor to account for the cosmic microwave background (CMB, see details further in the text), $D_{\rm L}$ is the luminosity distance in meter at redshift $z$, $B_{\nu_{\rm obs}}(T_{\rm obs})$ is Planck's blackbody function in Jy sr$^{-1}$ at the observed temperature $T_{\rm obs}$ ($T_{\rm obs}=T_{\rm rest}/(1+z)$), $\kappa_{\nu_0}$ is the photon cross section to mass ratio of dust in m$^2$\,kg$^{-1}$ and $\beta$ is the dust emissivity index. $S_{\nu_{obs}}$ is obtained by integrating over a fixed circular region in the center of the stack stamp (see Section \ref{sec:stacking}).  

Following \citet{Magnelli2020} and \citet{Pozzi2021} we decided to use a single cold component to account for the dust temperature, and used a mass weighted dust temperature of $T_{\rm rest}=25$\,K. While the total dust in galaxies is thought to consist of a warmer ($20<T<60$\,K) and a colder ($T<30$\,K) component, studies of local galaxies have shown that the cold component is responsible for most of the dust budget \citep[from 96\% to 99\%, see][]{Orellana2017} and for the majority of the Rayleigh-Jeans emission \citep[see also][for a study of the contribution of the cold and warm dust components to the dust continuum emission at $z\sim2$]{Shivaei2022}. \citet{Scoville2014, Scoville2016,Scoville2017} argue that this approximation should still be valid in higher redshift galaxies. We decided to adopt  $T_{\rm rest}=25$\,K at all redshifts probed by our analyses, but one should note that dust masses go as $T^{-1}$, meaning that our dust mass results are effectively highly dependent on the estimated dust temperature, and that the masses derived in our analyses could be off by factors of a few if the assumed dust temperature is not correct (see Section \ref{sec:discussion}). 

While choosing an evolving dust temperature with redshift could seem more appropriate, we decided to stick to a fixed value to allow easier comparison with similar previous studies. We used $\beta=1.8$, the Galactic dust emissivity index measurement from the {\it Planck} data \citep{Planck2011} which also correlates well with values observed in high-redshift galaxies \citep[e.g.][]{Chapin2009, Magnelli2012a, Faisst2020}. Theoretical studies predict values for $\beta$ ranging from 1.5 to 2.0 \citep[e.g.][]{Draine2011}, similarly observational studies such as \citet{Faisst2020} find $\beta$ values between 1.6 and 2.4 (with a median value of 2.0). The effect of $\beta$ is however less impactful than the choice of dust temperature and we find that choosing $\beta=1.5$ (or 2.0) would only modify our results by $\sim 9$\% at $z=1.5$ and $\sim 15$\% at $z=5.5$. We used $\kappa_{\nu_0}$=0.0431 m$^2$\,kg$^{-1}$ with ${\nu_0}=352.6$\,GHz \citep{Li2001,Magnelli2020}. See Section \ref{sec:discussion} for a more complete discussion of the choice of parameters for the dust mass computation. 

Following \citet{DaCunha2013} we corrected our flux measurements for the effect of CMB. First, we correct our dust temperature for the extra heating due to the CMB \citep[following Equation 12 of][]{DaCunha2013}. This effect is minor, as the dust temperature is only increased by $\sim 3\%$ at $z=6$ (the effect being even less important at lower redshift). Second, and most importantly, the CMB acts as a bright observing background, leading to an underestimation of the total flux. This is corrected by following Equation 18 of \citet{DaCunha2013}. This effect is much more important, as it yields an upward measurement to our results ranging from $\sim1.03$ at $z\sim1$ to $\sim1.31$ at $z\sim6$.

\subsection{Uncertainty computation} \label{sec:errors}

To compute the overall uncertainties associated to our measurements, we combined different sources of uncertainties. The first is the direct RMS from each stack analysis (computed from source-free stacks of each subsample, see Section \ref{sec:stacking}). In addition, in stacking analyses one should account for the intrinsic flux distribution of the sources in the sample. To do so, we perform a bootstrapping analysis for each subsample. In each subsample, the dust mass is re-computed 1000 times, each time with a different sample, randomized from the original subsample without replacement (see \citet{Jolly2020} for a detailed description of the bootstrapping routine included in \textsc{LineStacker}). The distribution of the dust masses computed in this manner are shown on Figures \ref{fig:boot_mass} and \ref{fig:boot_sfr}, where we also show the dust mass obtained from the original subsample (black vertical line). The peaks and shapes of the distributions are overall consistent with the values derived from the original samples. However, one can note that the peak of the distributions are, in most of the subsamples, slightly lower than the original stacks. Similarly, some distributions present a faint tail toward higher dust masses. These 2 informations combined hint for a skewed distribution, which can be easily explained from the presence of individual detections in the samples. These effects are small however, and should be well represented by the associated uncertainties on dust mass measurements. 

From the distributions derived from of each bootstrapping analyses we extract the 16$^{\mathrm{th}}$ to 84$^{\mathrm{th}}$ percentile (see Figures \ref{fig:boot_mass} and \ref{fig:boot_sfr}) that we sum quadratically with the stack RMS\footnote{One should note that the bootstrap analysis is also affected by the RMS of the stacked images. The two errors being hence correlated, one should in principle not add them quadratically. This is an approximation that should slightly over-estimate the error computation.}. In addition, given that the redshift is used multiple times in equation \ref{eq:dust}, its 16$^{\mathrm{th}}$ to 84$^{\mathrm{th}}$ percentile in each subsample is used to propagate the associated uncertainty on the dust mass computation. Finally, following \citet{Sun2022} and \citet{Fujimoto2023} we use a magnification uncertainty of 20\% of the magnification associated to each subsample, and propagate it to the aforementioned errors.

The $16^{\mathrm{th}}$ and $84^{\mathrm{th}}$ percentile of the (magnification corrected) SFRs and stellar masses in each subsample are used to compute the error associated to the SFRs and stellar masses.

The combination of these measurements uncertainties are used to plot the error bars on the different figures shown in this work. However, only the RMS associated to each cube is used to qualify a detection or an upper limit (as stated in section \ref{sec:stacking}).

\section{Results} \label{sec:results}

\subsection{Mean stacking results, dust continuum detection} \label{sec:res-stack}

\begin{figure*}
   \centering
    \includegraphics[width=0.9\linewidth]{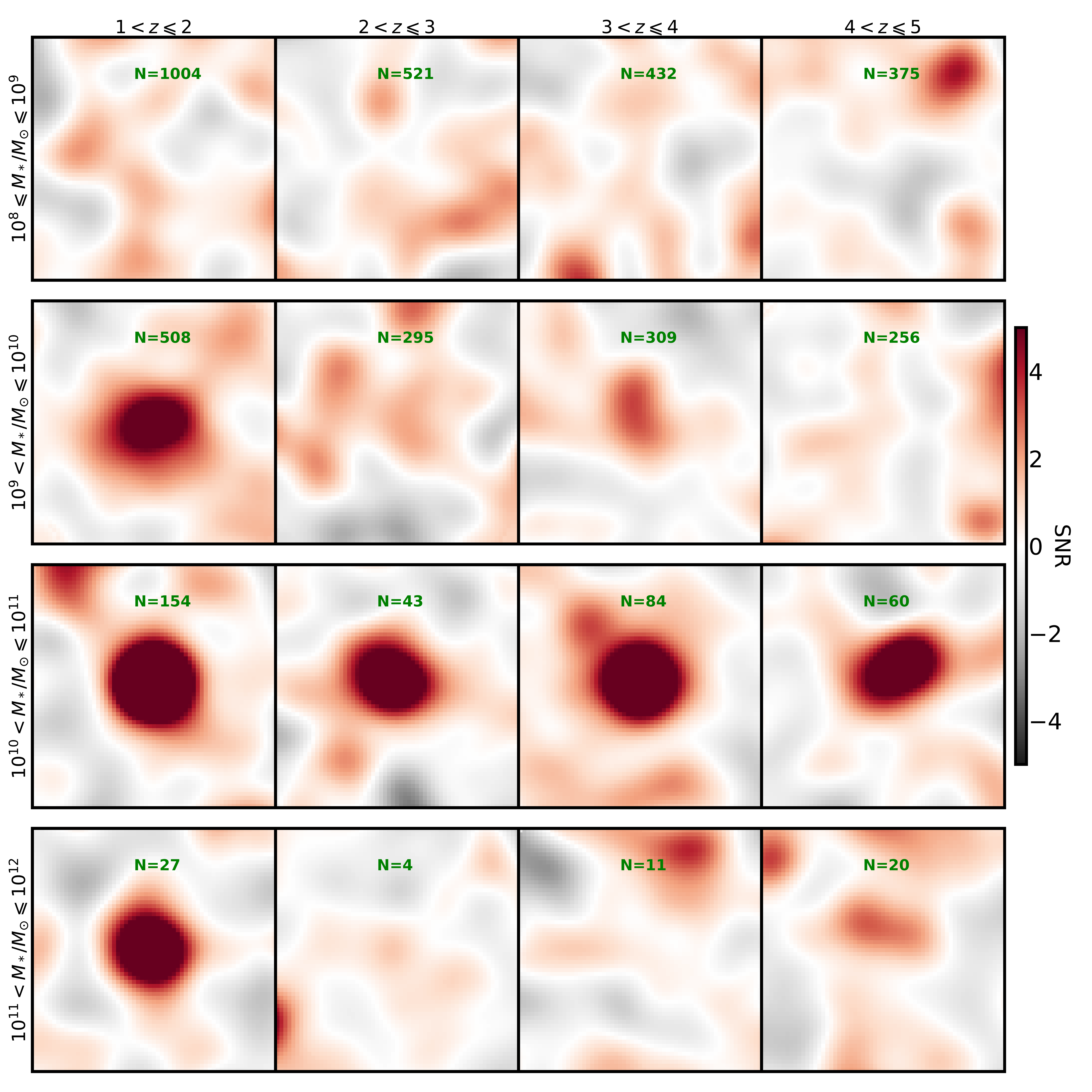}
      \caption{9.76 $\times$ 9.76 arcsec$^2$ (61 $\times$ 61 pixels) mean-stacking stamps, split in bins of stellar mass and redshift. Each map is normalised by the corresponding standard deviation computed in associated empty stacks (see Section \ref{sec:stacking}). The number of sources stacked (N) is indicated for each bin. }
         \label{fig:stamps_mstar}
\end{figure*}

\begin{figure*}
   \centering
    \includegraphics[width=0.9\linewidth]{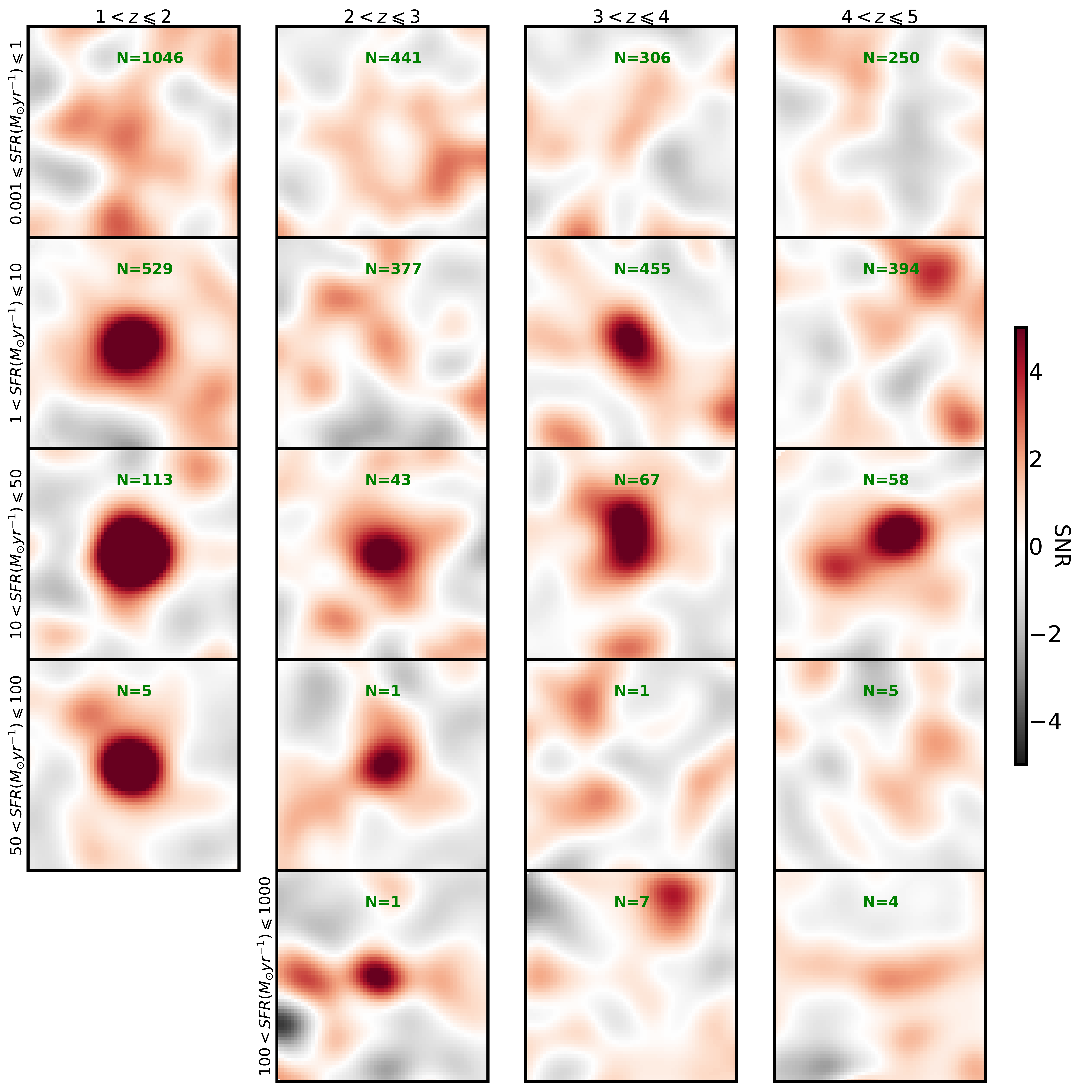}
      \caption{Similar to Figure \ref{fig:stamps_mstar} but using SFR bins instead of stellar mass ones.
              }
         \label{fig:stamps_sfr}
\end{figure*}

The standard deviation in stack stamps reaches levels as low as $\sim4.3 \times 10^{-3}$\,mJy per beam, in the $z=1-2$ bin, of the lowest stellar mass and SFR subsamples (both containing $\sim1000$ sources). The total flux in the central circular aperture of 2 arcsec radius varies between $\sim1.78 \times 10^{-2}$\,mJy and $\sim1.04$\,mJy for stack detections (flux above 3 $\sigma$, not corrected for magnification). Securely extracted dust masses (with S/N above 3\,$\sigma$) range from $\sim2.8 \times 10^6$\,M$_{\odot}$ to $\sim2.56 \times 10^8$\,M$_{\odot}$ (after magnification correction). Figures \ref{fig:stamps_mstar} and \ref{fig:stamps_sfr} show the stacked maps obtained for each of the subsamples, from which the results presented below are extracted. 

\begin{figure*}
   \centering
    \includegraphics[width=0.45\linewidth]{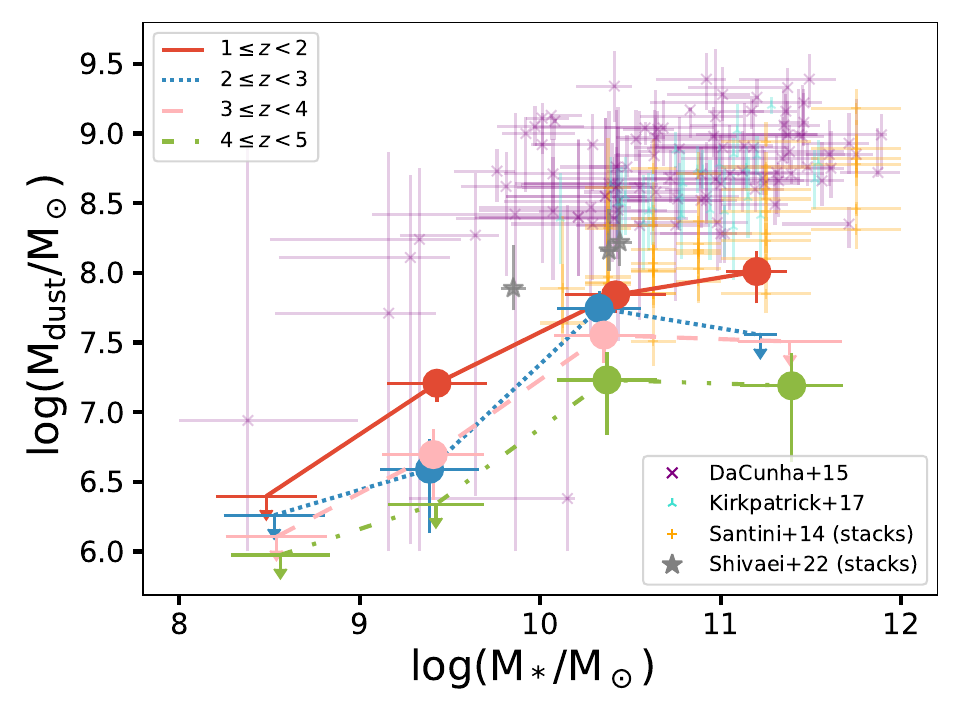}
    \includegraphics[width=0.45\linewidth]{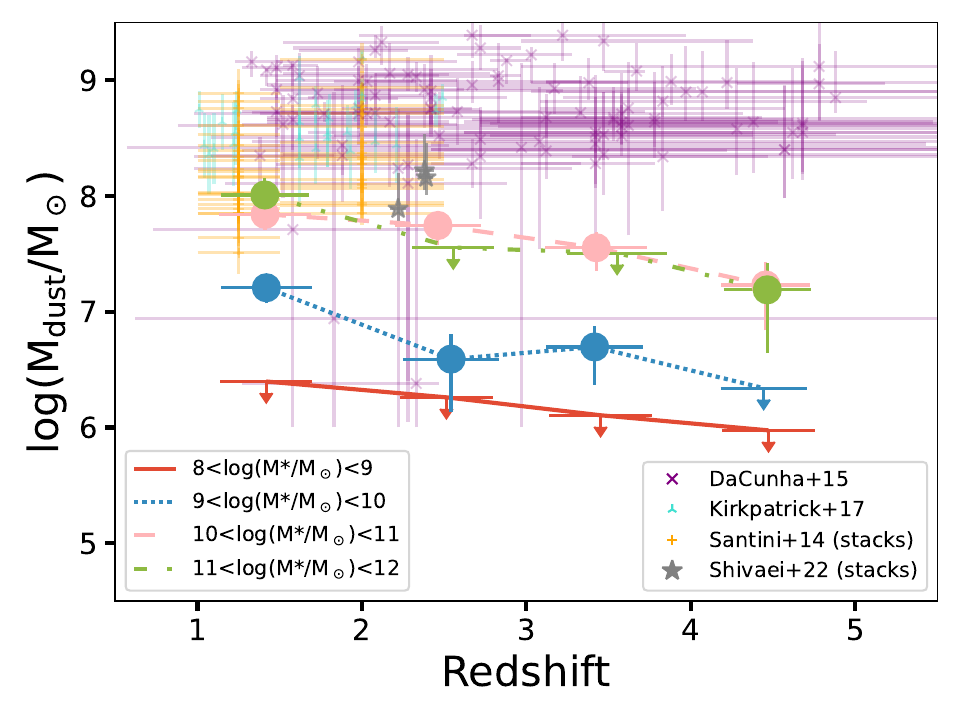}
      \caption{(left) Average dust mass as a function of the stellar mass in each redshift bin. Circles represent detections (above 3 $\sigma$) while down pointing arrows represent 3 $\sigma$ upper limits. Error bars on the dust mass detections are computed using a combination of different errors, see Section \ref{sec:errors}. Error bar along the x-axis represent the 16$^{\mathrm{th}}$ to 84$^{\mathrm{th}}$ percentile of parameter's distribution in the stacked sample. (right) Similar to the left panel but plotting as a function of redshift in each stellar mass bin. Also shown for comparison: individual dust mass measurements detections \citet{DaCunha2015,Kirkpatrick2017} and stacking measurements from \citet{Santini2014,Shivaei2022}.
              }
         \label{fig:mdust_mstar}
   \end{figure*}

\begin{figure}
   \centering
    \includegraphics[width=0.9\linewidth]{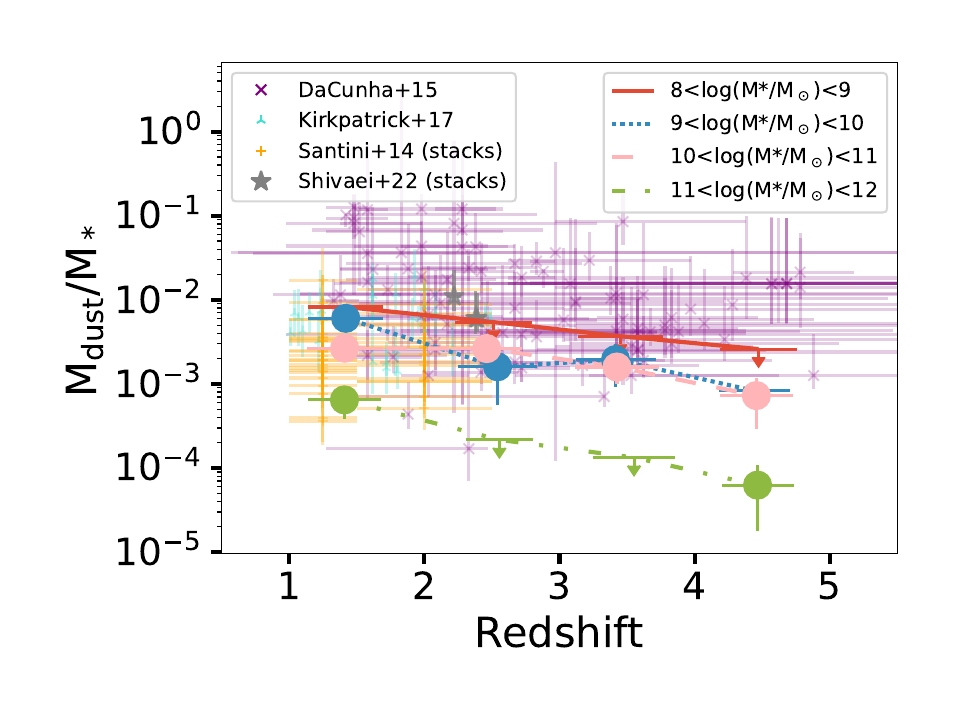}
      \caption{Average dust mass to average stellar mass ratio as a function of redshift. Similar to the right panel of Figure \ref{fig:mdust_mstar} but normalized by the stellar mass of galaxies in the sample. Error bars combine the error on dust mass computation and stellar mass distribution.}
         \label{fig:mdust_mstar_ratio}
   \end{figure}

\begin{figure*}
   \centering
    \includegraphics[width=0.45\linewidth]{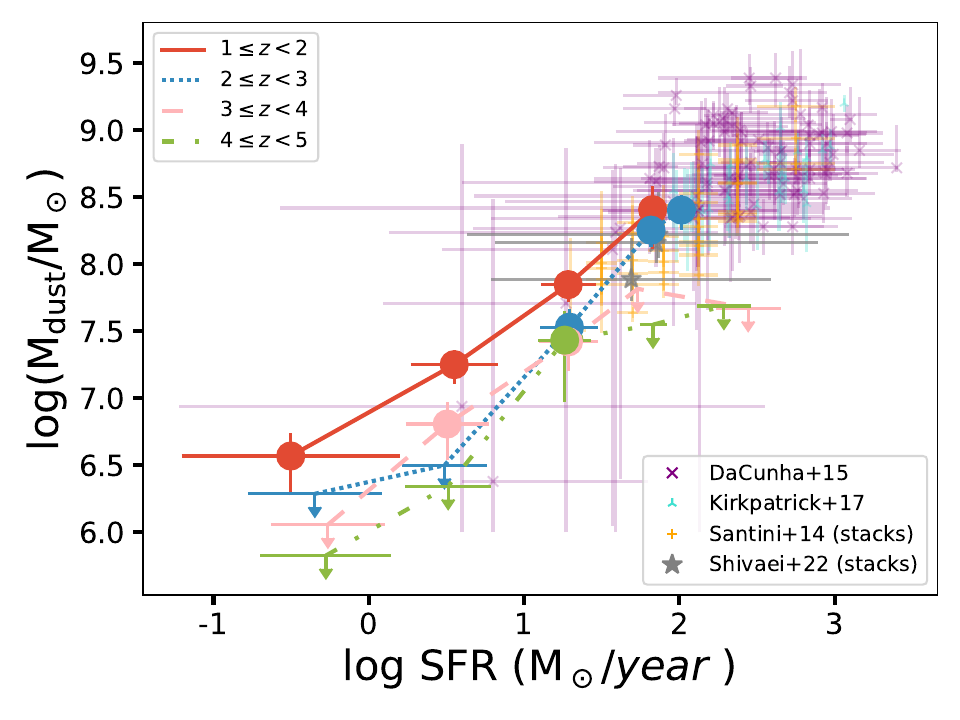}
    \includegraphics[width=0.45\linewidth]{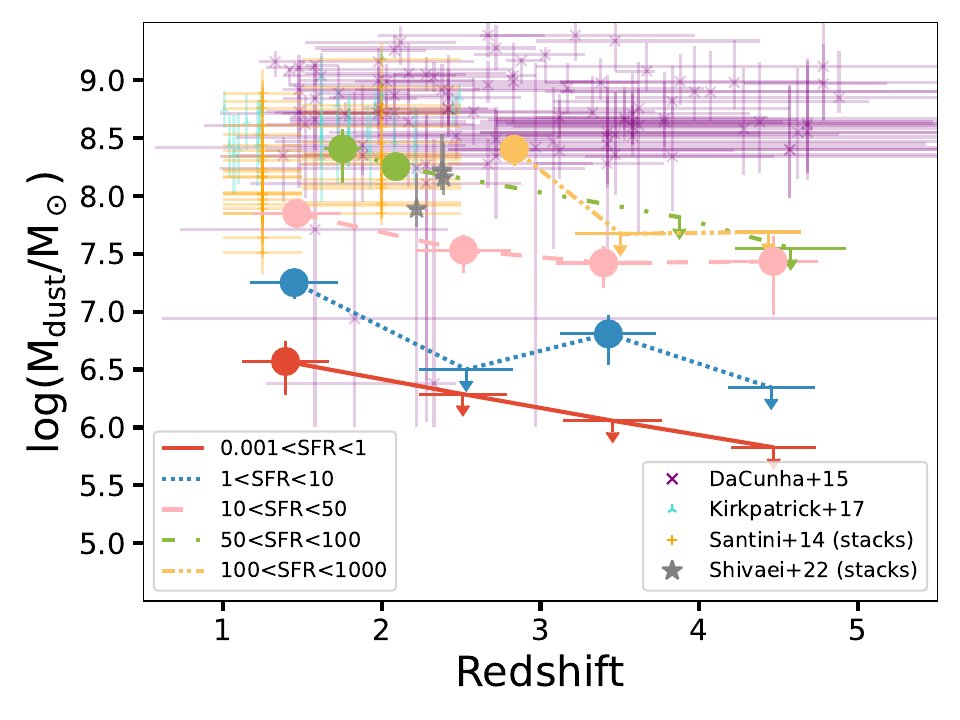}
      \caption{Similar to Figure \ref{fig:mdust_mstar} but in SFR bins instead of stellar mass.
              }
         \label{fig:mdust_sfr}
   \end{figure*}

The evolution of dust mass with stellar mass and redshift is shown on Figure \ref{fig:mdust_mstar}, and tabulated on Table \ref{tab:stack_res_mstar}. While detections in the lowest redshift bin ($1<z<2$) indicate a clear linear relationship between log(M$_{\mathrm{dust}}$) and log(M$_*$), the other redshift bins are less regular (see left panel of Figure \ref{fig:mdust_mstar}). When comparing to other works \citep[][also plotted on the figures]{Santini2014,DaCunha2015,Kirkpatrick2017,Shivaei2022}, our data points are typically below other measurements (individual detections and stacks). This behavior is however strongly reduced when excluding quiescent galaxies from the stacks (see Appendix), highlighting that quiescent galaxies might have lower dust masses at the same stellar mass.

The log($M_{\mathrm{dust}}$)\,--\,log(SFR) relation seems to follow a similar trend of increasing dust mass with increasing SFR, but with a clearer linear relation across redshift (though there is a higher number of non-detection, see left panel of Figure \ref{fig:mdust_sfr} and stacking results tabulated in Table \ref{tab:stack_res_sfr}). Hinting that SFR might be a better suited than stellar mass to trace the dust mass. When comparing to previous works \citep{Santini2014,DaCunha2015,Kirkpatrick2017,Shivaei2022} one can see that our measurements follow a similar trend as the one already observed, reaching however so far mostly unexplored regimes.

\begin{table*}
\caption{\label{tab:stack_res_mstar} Stacking results in stellar mass subsamples. When the central flux is lower than 3 times the rms of the corresponding stack map: the 3 $\sigma$ upper limit is shown in place of the dust mass. Errors on the computed dust mass correspond to the quadratic sum of the standard deviation computed from empty stacks with the uncertainties coming from the bootstrap analysis, and to which is finally propagated the magnification error (see Section \ref{sec:errors}). Flux and RMS are extracted as indicated in Section \ref{sec:stacking}: flux is integrated in the central circular region of the stack stamp, with a radius $R_{\mathrm{integ}}=2''$, and then converted to Jy from Jy/beam. RMS is computed from the standard deviation in empty stacks. $(a)$ <$\mu$> is the average magnification of the sources in the subsample.}
\centering   
 \begin{tabular}{cccccc}
 \hline \hline
 <$z$> & log$($<$|M_*|$>$/M_\odot)$ & <Dust Mass> & Flux & RMS & <$\mu$>$^{(a)}$ \\  
  &  & (\msun) & (mJy) & (mJy) & \\
\hline
$1.42$ & $8.48$ & <$2.49\times10^{6}$ & -- & $4.28\times10^{-3}$  & $3.43$  \\ 
$2.51$ & $8.53$ & <$1.82\times10^{6}$ & -- & $5.64\times10^{-3}$  & $5.41$  \\ 
$3.45$ & $8.54$ & <$1.27\times10^{6}$ & -- & $6.51\times10^{-3}$  & $7.83$  \\ 
$4.47$ & $8.56$ & <$9.43\times10^{5}$ & -- & $6.98\times10^{-3}$  & $10.55$  \\ 
[0.3cm] 
 \hline 
$1.42$ & $9.43$ & $1.61\times10^{7}$ $\pm$ $4.33\times10^{6}$ & $7.84\times10^{-2}$ & $6.67\times10^{-3}$  & $3.24$  \\ 
$2.54$ & $9.39$ & $3.89\times10^{6}$ $\pm$ $2.54\times10^{6}$ & $2.54\times10^{-2}$ & $8.25\times10^{-3}$  & $3.78$  \\ 
$3.41$ & $9.41$ & $4.96\times10^{6}$ $\pm$ $2.61\times10^{6}$ & $4.49\times10^{-2}$ & $9.12\times10^{-3}$  & $4.64$  \\ 
$4.44$ & $9.42$ & <$2.19\times10^{6}$ & -- & $9.12\times10^{-3}$  & $5.95$  \\ 
[0.3cm] 
 \hline 
$1.41$ & $10.42$ & $6.95\times10^{7}$ $\pm$ $1.84\times10^{7}$ & $0.28$ & $1.27\times10^{-2}$  & $2.67$  \\ 
$2.46$ & $10.33$ & $5.57\times10^{7}$ $\pm$ $1.78\times10^{7}$ & $0.3$ & $2.37\times10^{-2}$  & $3.16$  \\ 
$3.43$ & $10.35$ & $3.57\times10^{7}$ $\pm$ $1.32\times10^{7}$ & $0.33$ & $1.96\times10^{-2}$  & $4.71$  \\ 
$4.46$ & $10.37$ & $1.7\times10^{7}$ $\pm$ $1.01\times10^{7}$ & $0.2$ & $2.02\times10^{-2}$  & $5.67$  \\ 
[0.3cm] 
 \hline 
$1.41$ & $11.2$ & $1.02\times10^{8}$ $\pm$ $4.17\times10^{7}$ & $0.39$ & $2.88\times10^{-2}$  & $2.53$  \\ 
$2.56$ & $11.22$ & <$3.58\times10^{7}$ & -- & $9.51\times10^{-2}$  & $4.6$  \\ 
$3.55$ & $11.38$ & <$3.21\times10^{7}$ & -- & $4.93\times10^{-2}$  & $2.33$  \\ 
$4.46$ & $11.39$ & $1.55\times10^{7}$ $\pm$ $1.11\times10^{7}$ & $0.16$ & $4.03\times10^{-2}$  & $4.87$  \\ 
[0.3cm] 
 \hline 
 \end{tabular}
\end{table*}

\begin{table*}
\caption{\label{tab:stack_res_sfr} Similar to Table \ref{tab:stack_res_mstar} but for SFR subsamples.}
\centering   
 \begin{tabular}{cccccc}
 \hline \hline
 <$z$> & <SFR> & <Dust Mass> & Flux & RMS & <$\mu$> \\  
  & (\msun / yr) & (\msun) & (mJy) & (mJy) & \\
\hline
$1.39$ & $0.15$ & $3.72\times10^{6}$ $\pm$ $1.82\times10^{6}$ & $1.95\times10^{-2}$ & $4.28\times10^{-3}$  & $3.51$  \\ 
$2.51$ & $0.81$ & <$1.94\times10^{6}$ & -- & $6.17\times10^{-3}$  & $5.55$  \\ 
$3.45$ & $0.99$ & <$1.14\times10^{6}$ & -- & $7.17\times10^{-3}$  & $9.62$  \\ 
$4.47$ & $0.53$ & <$6.72\times10^{5}$ & -- & $7.05\times10^{-3}$  & $14.94$  \\ 
[0.3cm] 
 \hline 
$1.45$ & $2.84$ & $1.78\times10^{7}$ $\pm$ $4.96\times10^{6}$ & $8.13\times10^{-2}$ & $6.47\times10^{-3}$  & $3.04$  \\ 
$2.53$ & $1.79$ & <$3.17\times10^{6}$ & -- & $7.19\times10^{-3}$  & $3.95$  \\ 
$3.43$ & $2.16$ & $6.43\times10^{6}$ $\pm$ $2.94\times10^{6}$ & $5.51\times10^{-2}$ & $7.36\times10^{-3}$  & $4.38$  \\ 
$4.45$ & $4.68$ & <$2.21\times10^{6}$ & -- & $7.72\times10^{-3}$  & $4.97$  \\ 
[0.3cm] 
 \hline 
$1.47$ & $24.56$ & $7.06\times10^{7}$ $\pm$ $1.83\times10^{7}$ & $0.26$ & $1.38\times10^{-2}$  & $2.5$  \\ 
$2.52$ & $19.66$ & $3.4\times10^{7}$ $\pm$ $1.23\times10^{7}$ & $0.2$ & $2.29\times10^{-2}$  & $3.49$  \\ 
$3.4$ & $31.67$ & $2.64\times10^{7}$ $\pm$ $1.05\times10^{7}$ & $0.22$ & $2.37\times10^{-2}$  & $4.23$  \\ 
$4.46$ & $10.55$ & $2.72\times10^{7}$ $\pm$ $1.78\times10^{7}$ & $0.19$ & $2.29\times10^{-2}$  & $3.4$  \\ 
[0.3cm] 
 \hline 
$1.76$ & $86.71$ & $2.56\times10^{8}$ $\pm$ $1.26\times10^{8}$ & $1.04$ & $7.88\times10^{-2}$  & $2.64$  \\ 
$2.09$ & $60.65$ & $1.8\times10^{8}$ $\pm$ $4.46\times10^{7}$ & $0.65$ & $9.54\times10^{-2}$  & $2.25$  \\ 
$3.87$ & $81.75$ & <$6.56\times10^{7}$ & -- & $0.1$  & $2.35$  \\ 
$4.58$ & $53.34$ & <$3.55\times10^{7}$ & -- & $6.11\times10^{-2}$  & $2.45$  \\ 
[0.3cm] 
 \hline 
$2.84$ & $1.33\times10^{2}$ & $2.54\times10^{8}$ $\pm$ $7.27\times10^{7}$ & $1.0$ & $0.2$  & $2.17$  \\ 
$3.5$ & $1.51\times10^{2}$ & <$4.71\times10^{7}$ & -- & $6.67\times10^{-2}$  & $2.16$  \\ 
$4.44$ & $3.52\times10^{2}$ & <$4.89\times10^{7}$ & -- & $0.1$  & $2.93$  \\ 
[0.3cm] 
 \hline 
 \end{tabular}
\end{table*}

When looking at the evolution of dust mass with redshift (see the right panels of Figures \ref{fig:mdust_mstar} and \ref{fig:mdust_sfr} and Figure \ref{fig:mdust_mstar_ratio}), one can see a relatively clear linear trend, indicating a decrease in dust mass with increasing redshift. 

By fitting quadratic functions to the data shown on Figures \ref{fig:mdust_mstar} and \ref{fig:mdust_sfr}, we derive scaling relations between: $M_{\mathrm{dust}}$ and $M_*$ at fixed $z$; $M_{\mathrm{dust}}$ and $z$ at fixed $M_*$; $M_{\mathrm{dust}}$ and SFR at fixed $z$; and $M_{\mathrm{dust}}$ and $z$ at fixed SFR. Fits are performed using detections only (ignoring upper limits), and only when the number of data points is $\geq 3$. The results from the fits are summarized in Table \ref{tab:fit}, and plotted on Figures \ref{fig:mdust_mass_fit} and \ref{fig:mdust_sfr_fit}

\begin{table*}
\caption{\label{tab:fit} Fitting results. Fits are performed with function of the form $M_{\mathrm{dust}}(x)=a x^b$ where $a$ and $b$ are free parameters and $x$ is respectively the SFR, $z$ or $M_*$. The range of applicability refers to the parameter range over which subsamples yielded detections. See Figures \ref{fig:mdust_sfr_fit} and \ref{fig:mdust_mass_fit} to visualize the fits.}
\centering   
 \begin{tabular}{lccc}
 \hline \hline
$M_{\mathrm{dust}}$(SFR) (fixed $z$)\\
$z$ range & a & b & Range of applicability\\ 
\hline
$1\leq z<2$ & $7.2 \pm 1.2 \times 10^{6}$ & $7.8 \pm 0.7 \times 10^{-1}$ & $-1<$log(SFR)$<2$\\
$2\leq z<3$ & $9.2 \pm 5.7 \times 10^{5}$ & $12.3 \pm 1.5 \times 10^{-1}$ & $1<$log(SFR)$<3$ \\
[0.3cm]

$M_{\mathrm{dust}}(z)$ (fixed SFR)\\
SFR range ($M_\odot$ per year) & a & b & Range of applicability \\ 
\hline
$10<\mathrm{SFR}<50 $ & $10.5 \pm 1.6 \times 10^{7}$ & $-1.1 \pm 0.2$ & $1\leq z<5$ \\
[0.3cm]

$M_{\mathrm{dust}}(M_*)$ (fixed $z$)\\
$z$ range & a & b & Range of applicability \\ 
\hline
$1\leq z<2$ & $5.2 \pm 13.0 \times 10^{3}$ &  $4.8 \pm 1.1 \times 10^{-1}$ & $8<$log(M$_*$/M$_\odot$)$<12$ \\
[0.3cm]

$M_{\mathrm{dust}}(z)$ (fixed $M_*$)\\
$M_*$ range ($M_\odot$) & a & b & Range of applicability \\ 
\hline
$10^9<\mathrm{M}_*<10^{10}$ & $3.0 \pm 1.3 \times 10^{7}$ & $-1.8 \pm 0.7$ & $1\leq z<4$ \\
$10^{10}<\mathrm{M}_*<10^{11}$ & $10.5 \pm 2.7 \times 10^{7}$ & $-1.0 \pm 0.3$ & $1\leq z<4$ \\
\hline
\end{tabular}

\end{table*}

\subsection{Median stacking analysis} \label{sec:median_stacking}

\begin{figure*}
   \centering
    \includegraphics[width=0.45\linewidth]{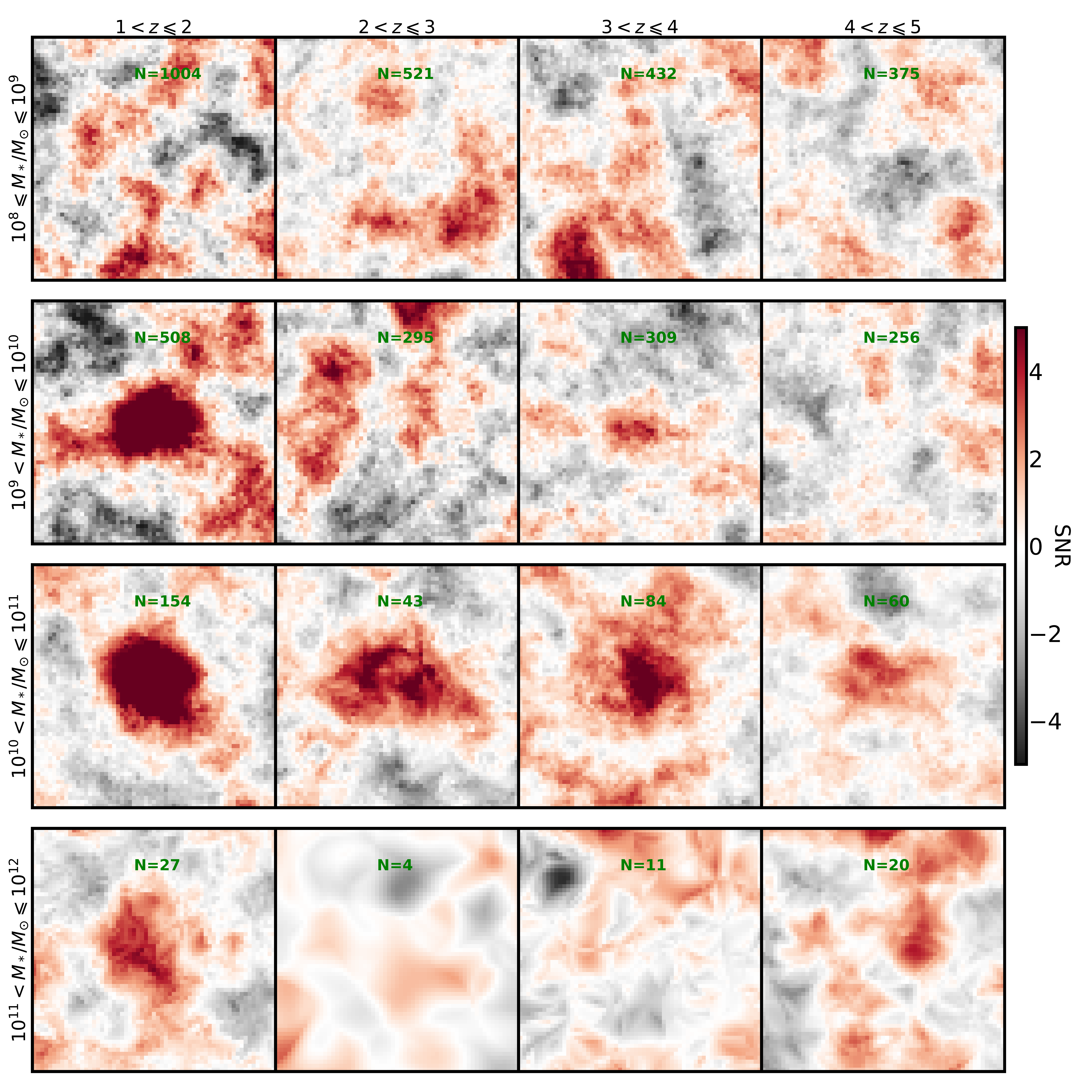}
    \includegraphics[width=0.45\linewidth]{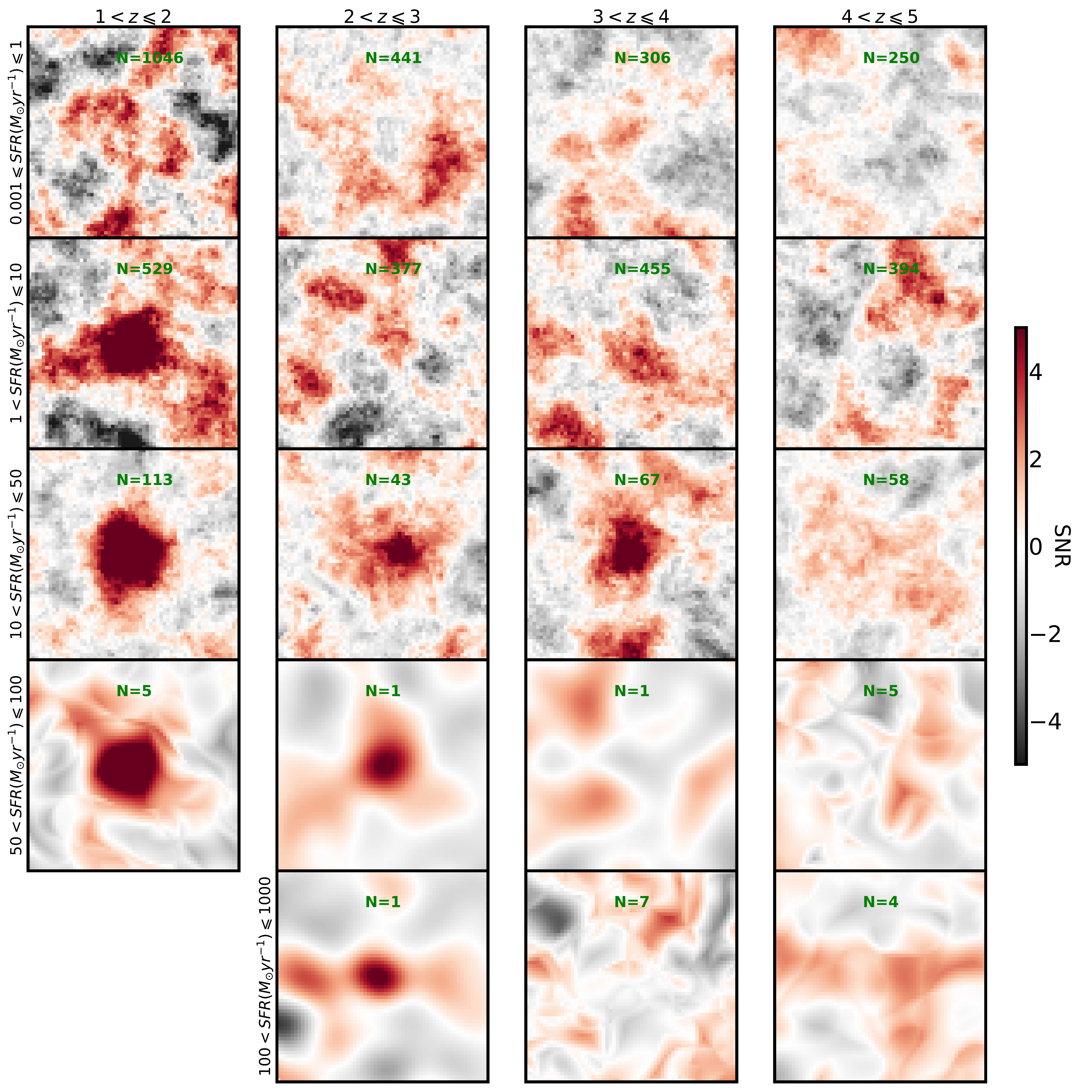}
    
      \caption{Median stacking stamps, split in stellar mass and redshift (left panel) and SFR and redshift (right panel). Similar to Figures
      \ref{fig:stamps_mstar} and \ref{fig:stamps_sfr}.
              }
         \label{fig:stamps_median}
   \end{figure*}

To better assess the effect of the population distribution on the derived dust mass we also performed median-stacks, in the same way as the average stacks presented in Section \ref{sec:stacking}. The overall median dust masses derived through our stacking analyses are systematically slightly lower than the mean dust masses (see Figure \ref{fig:stamps_median} for illustration and the appendix for detailed plots). While the median stellar mass and SFR of each subsample are also systematically lower than the mean (see Figures \ref{fig:mass_subsample_hist} and \ref{fig:sfr_subsample_hist}), the difference is not big enough to explain the offset dust mass observed between the mean and median stacks. This, again, hints to a possibly skewed distribution of dust mass in each subsample, resulting in offsets between mean and median dust masses. However, it should be however noted that the dust-mass trends observed in this analysis (with redshift, stellar mass and SFR) remain the same in the median stacking analyses.

\section{Discussion} \label{sec:discussion}

As our analysis focuses on galaxies spanning a large redshift range, and our observation remains at a fixed observed wavelength, it is important to question the assumption made here; that the continuum flux traces the same physical process at a rest wavelength $\lambda_{\mathrm{rest}}\sim0.5$\,mm (corresponding to $z\sim1$) and $\lambda_{\mathrm{rest}}\sim0.18$\,mm (corresponding to $z\sim5$). More specifically, the validity of the dust mass equation (see Section \ref{sec:dust}) comes from the assumption that the dust is optically thin, and that the flux measurement probes the Rayleigh-Jeans dust emission --thus the part of the SED that is dominated by the emission of the dust at T=25K, that dominates the dust mass budget. This is true at long wavelengths, where we effectively probe the Rayleigh-Jeans tail of the dust emission \citep[e.g.][]{Magdis2012, Scoville2016}. However, the higher the redshift observed the closer one gets to the peak of dust emission, more sensitive to temperature and total luminosity. This could explain the low detection rate in the high-$z$ subsamples, even when the number of stack sources is high --and hence the corresponding RMS in the stack maps is low.

While numerous studies show an evolution of dust temperature with redshift, it is important to distinguish between the peak temperature, which is thought from both theoretical and observational works to increase with redshift \citep[e.g. ][]{Magdis2012,Magnelli2014,Bethermin2015,Ferrara2017,Narayanan2018,Schreiber2018,Liang2019,Ma2019,Sommovigo2020}, and the mass-weighted temperature, probing the Rayleigh-Jeans tail, which can be directly used to measure the dust mass as stated above. Unlike the peak temperature, the mass-weighted temperature is thought to be mostly constant with redshift \citep[see][]{Scoville2016,Liang2019}. It should be noted however, that the mass-weighted temperature still exhibits a small range in possible values, typically from 15 to 45\,K \citep[e.g.][]{Liang2019}, and a small increase with redshift may exist. On the other hand, \citet{Dudzeviciute2020} argue that the previously observed evolution of the dust temperature is likely due to the luminosity evolution in the samples employed, and suggest a dust temperature of $30.4 \pm 0.3$, constant with redshift. In any event, to ease comparison with studies similar to ours and to limit complexity, we decided to keep a fixed temperature at all redshifts for dust mass computations. Consequently, one should note that using $T=40$\,K (instead of 25) in dust masses computation would change our results by $\sim$\,50\% in the highest redshift bin.

\begin{figure}
   \centering
        \includegraphics[width=0.9\linewidth]{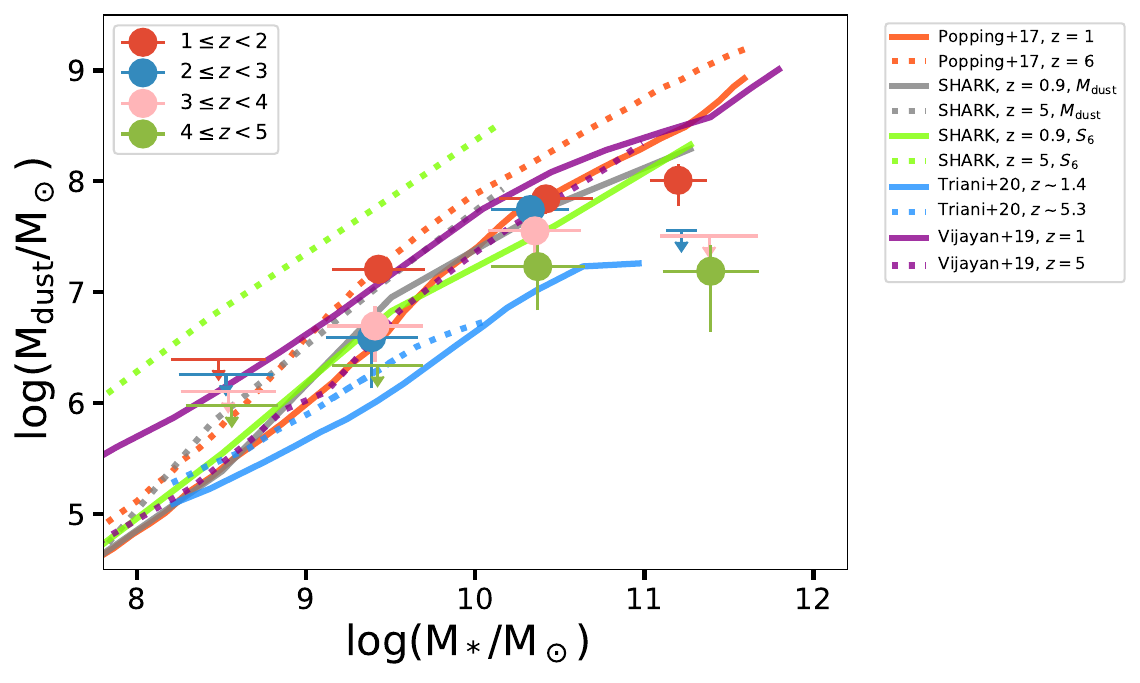}
      \caption{Log average dust mass recovered as a function of the (log) stellar mass, in each redshift bin. Circles represent detections (above 3 $\sigma$) while 3$\sigma$ upper limits are represented by down pointing arrows. Overplotted, the $z=1$ and $z=6$ dust mass-stellar mass relation from \citet{Popping2017} and the $z\sim1$ and $z\sim5$ relations from \citet{Shark} (SHARK model), \cite{Triani2020}, \cite{Vijayan2019}. Both the average dust mass directly predicted by the SHARK model, and the dust mass computed from the predicted band 6 flux --following equation \ref{eq:dust}-- are shown.}
         \label{fig:mdust_mstar_simu}
\end{figure}

\begin{figure}
   \centering
    \includegraphics[width=0.9\linewidth]{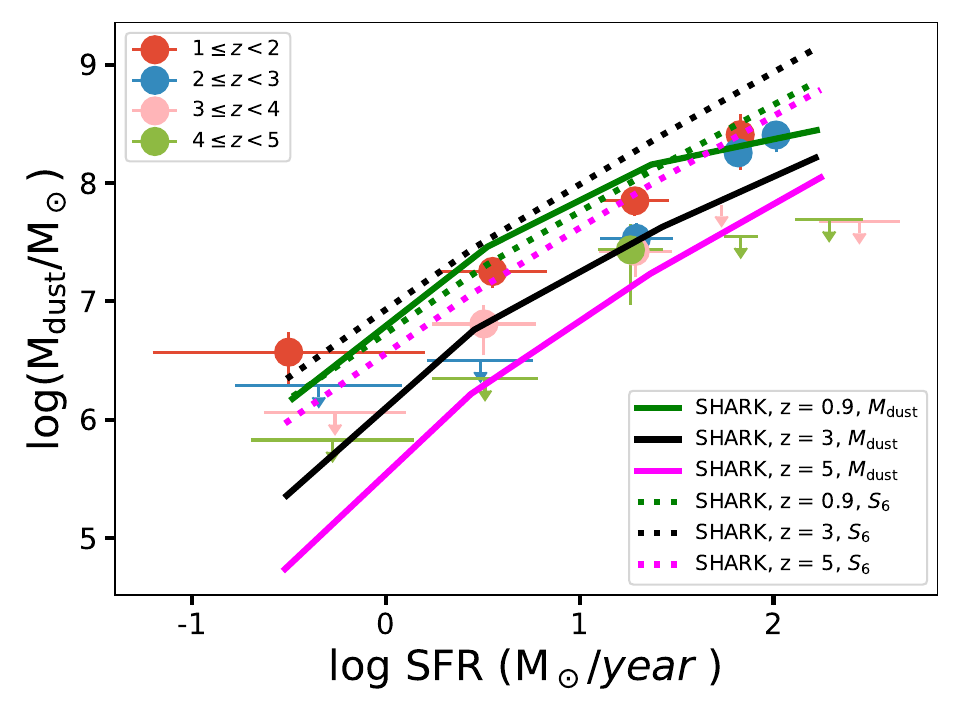}
      \caption{Similar to Figure \ref{fig:mdust_mstar_simu} but for SFR instead of stellar mass. Prediction from \citet[][SHARK model]{Shark} at $z\sim1$, $3$ and $5$ are shown, both in terms of direct dust mass predictions and dust mass inferred from predicted band 6 flux --following equation \ref{eq:dust}.}
         \label{fig:mdust_sfr_simu}
\end{figure}

When comparing our results to the semi-analytical modeling from \citet[][see Figures \ref{fig:mdust_mstar_simu} and \ref{fig:mdust_sfr_simu}]{Popping2017,Shark,Vijayan2019,Triani2020}\footnote{One should note that the SHARK model \citep{Shark} does not directly track the buildup and destruction of dust. Instead, dust masses are estimated assuming an empirical $z=0$ relation between dust-to-gas ratio and gas-phase metallicity.} we can see an overall decent agreement in dust mass predictions as a function of stellar mass at $z\sim1$. However, all models but the one presented in \citet{Vijayan2019} predict an overall increase of dust mass with redshift while our analysis shows the opposite. In addition the dust mass derived in the high-mass end of our subsamples is typically smaller than predicted by the models (though this flattening might be due to the presence of quiescent galaxies in our sample, as shown in the appendix). In addition, one should note that the method used for computing the average dust mass in the SHARK model \citep{Shark} differs to the one used in this work. To highlight this difference we show, on Figures \ref{fig:mdust_mstar_simu} and \ref{fig:mdust_sfr_simu}, both the average dust masses predicted by the SHARK model and the dust masses computed, using equation \ref{eq:dust}, from the band 6 continuum (S6) predicted by SHARK. It is interesting to note that at $z\sim3$ and at $z\sim5$ the dust mass derived using the predicted S6 differs by almost one dex from the dust mass otherwise predicted. As mentioned above, this might be a direct consequence of the evolution of the rest-wavelength with redshift, moving observation closer to the peak of dust emission. Surprisingly however, our results seem in better agreement with the direct dust mass measurements compared to the ones obtained from S6.

As the analysis relies on photometric redshifts for a large fraction of the catalog, it is important to consider the potential impact this could have on the results. In particular, it should be noted that the bins at $z>3$ appear overpopulated. This can be seen on Figure \ref{fig:SMF}, when comparing the SMF of our sample to the SMF of the COSMOS2020 sample \cite{Weaver2022,Weaver2023}. As a consequence the dust content at $z>3$ is likely over-estimated. This is probably due to the miss-classification of some lower redshift sources to higher redshift, as suggested in \citet{Kokorev2022} and mentioned in Section \ref{sec:cat}.

It should also be noted that, because the catalog used was generated using {\it HST} and \textit{Spitzer} photometry, some of the most dusty galaxies may be missing, biasing the analysis toward lower averaged dust masses. Indeed, as pointed out in \citet{Kokorev2022}, from the 180 sources individually detected with a SNR $>4$ in the ALCS data, only 145 are identified in their catalog, indicating that some of the most dusty galaxies may be missed from our analysis. This bias could be strongest in the high-redshift bins, where only the UV-brightest galaxies (i.e. typically less dust-obscured) might be observed. The use of James-Webb Space Telescope (JWST)-based catalogs could help to include more of the missed dusty galaxies, especially at high-$z$, and to improve the SED fitting routines.

As shown in the appendix (Figures \ref{fig:mdust_mass_dMS} and \ref{fig:mdust_sfr_dMS}), the inclusion of quiescent galaxies in the stacks biases the analysis toward lower dust masses, especially in the samples split through stellar-mass. This highlights that some sources included in the high-mass subsamples have low SFR and also low dust masses. The higher average dust mass observed in the mass-selected samples when excluding sources with $\Delta\mathrm{MS}<-0.5$, and on the other hand, the relatively stable average dust mass in SFR selected samples, implies that SFR should be a better dust mass tracer. It shows in addition that quiescent galaxies will have a typically lower dust content than their star forming counterparts at similar stellar mass.   

As in \cite{Magnelli2020}, we use the dust opacities from \cite{Li2001} to derive the dust masses. Those values have been constrained from observations of the diffuse interstellar medium (ISM) and therefore these properties are representative for dust in the diffuse ISM. However, some studies in nearby galaxies \citep[e.g.][]{galliano_interstellar_2018} showed that the opacity from \cite{Li2001} has perhaps been underestimated by a factor of $2-3$. One must emphasise that larger values of opacities in galaxies does not mean that the values of \cite{Li2001} are wrong. Their values have been constrained for the diffuse interstellar medium and are therefore well adapted for galaxies only if the bulk of dust emission emerges from the diffuse interstellar medium of galaxies. The underestimation of the dust opacity might indicate that the majority of the dust emission does not emerge from diffuse regions but from denser regions where dust grains are likely to be larger.  

In dense regions of the ISM, dust grains grow through accretion and coagulation which triggers an increase in the dust opacity by a factor $2-3$ \citep[e.g.][]{kohler_dust_2015}. The underestimation of the dust opacities in some nearby galaxies could therefore mean that the bulk of the dust emission arises from dense regions where dust grains are larger than in the diffuse ISM. This seems to be also the case in high-redshift galaxies as \cite{Magnelli2020} claim that the bulk of the dust emission in their galaxy sample emerges from the molecular phase and therefore dense regions where dust differs from the diffuse ISM. They also stated that there could be a significant increase in the dust emissivity from diffuse to denser regions which can be explained in terms of grain growth \citep[e.g.][]{kohler_dust_2015}. 

There is for now no consensus on the dust properties one must consider, neither in nearby nor in high-redshift galaxies. Moreover, dust grains are unlikely to be the same within galaxies and from one galaxy to another, which adds to the complexity. Although diffuse ISM-like dust grains might not be responsible for the majority of the dust emission in galaxies, these dust models have the advantage of being thoroughly developed and extensively used. We therefore use those dust properties from the diffuse ISM until we have more constraints on the dust grains responsible for the emission in these galaxies.

\section{Summary} \label{sec:summary}

We used deep ALMA band 6 data, over 33 lensing clusters, to perform a set of stacking analyses on a large sample of 4103 lensed sources, whose positions and physical parameters were extracted from the HST/IRAC catalog presented in \citet{Kokorev2022}. Spanning redshifts from $z=1$ to $z=5$, SFRs from $0.001 \, \mathrm{M}_{\odot} \, \mathrm{yr}^{-1}$ to $1000 \, \mathrm{M}_{\odot}\,\mathrm{yr}^{-1}$ and stellar masses from $10^{8}\, \mathrm{M}_{\odot}$ to $10^{12}\, \mathrm{M}_{\odot}$. To study the evolution of the dust mass through cosmic time, we binned the sources in five different redshifts bins, and further also grouped sources according to their SFR and stellar mass, respectively. We performed continuum stacking on each of the subsamples using \textsc{LineStacker} \citep{Jolly2020}. 

Using the mean continuum stacked flux, we computed average dust masses. From the non detections we derive $3\sigma$ upper limits. Using both detections and upper limits we study the evolution trend of dust mass with redshift. We see clear indications for average dust mass diminution with redshift. Similarly we study the evolution of dust mass with stellar mass and SFR, in both cases we see a positive correlation. From our detections we derived relationships between dust mass and SFR or stellar mass in different redshift range. The results from our study are mostly consistent with results from modeling (at least at $z\sim1$) and from other similar studies. Our analysis allows us to probe regimes of stellar mass and SFR unexplored so far. The highest redshift bin shows mostly non-detections, and the corresponding upper limits indicate low average dust masses --even though the overall RMS is low-- when compared to other studies. This could be due to the different rest-wavelength probed at high-$z$, lying closer to the peak of dust emission. Or to the tendency of individual high-$z$ dust measurements to be biased towards very dust-bright objects.  

The linear trends observed between stellar mass and dust mass confirm that as galaxies evolve and form more stars, they also accumulate more dust. These dust grains play a vital role in catalyzing the formation of H$_2$, thus influencing the overall gas reservoir available for future star formation. Dust becomes an integral component of a self-perpetuating cycle, accumulating throughout the stars' life cycle and subsequently bolstering the SFR. In addition the inverse trend observed between average dust-mass and redshift, implies that the build-up of dust in galaxies over cosmic-time is a gradual process, aligning with the overall evolution of the SFR density \citep[e.g][]{Magnelli2020}. This further hints for a gradual increase of the average-metallicity in galaxies with cosmic time, as dust grains are known to be composed of heavy-elements produced in stars.

\begin{acknowledgements}
The authors thank the anonymous reviewer for their thoughtful and constructive suggestions, which greatly helped improve this manuscript. This paper makes use of the ALMA data:  ALMA \#2018.1.00035.L, \#2013.1.00999.S, and \#2015.1.01425.S. ALMA is a partnership of the ESO (representing its member states), NSF (USA) and NINS (Japan), together with NRC (Canada), MOST and ASIAA (Taiwan), and KASI (Republic of Korea), in cooperation with the Republic of Chile. The Joint ALMA Observatory is operated by the ESO, AUI/NRAO, and NAOJ. JBJ thanks Ian Smail for the discussion. KK acknowledges support from the Swedish Research Council (2015-05580), and the Knut and Alice Wallenberg Foundation (KAW 2020.0081). K. Kohno acknowledges the JSPS KAKENHI Grant Number JP17H06130 and the NAOJ ALMA Scientific Research Grant Number 2017-06B. AG acknowledges funding from ANID-Chile NCN2023\_002 and FONDECYT Regular 1171506. DE acknowledges support from a Beatriz Galindo senior fellowship (BG20/00224) from the Spanish Ministry of Science and Innovation, projects PID2020-114414GB-100 and PID2020-113689GB-I00 financed by MCIN/AEI/10.13039/501100011033, project P20\_00334  financed by the Junta de Andaluc\'{i}a, project A-FQM-510-UGR20 of the FEDER/Junta de Andaluc\'{i}a-Consejer\'{i}a de Transformaci\'{o}n Econ\'{o}mica, Industria, Conocimiento y Universidades. FEB acknowledges support from ANID-Chile BASAL CATA FB210003, FONDECYT Regular 1200495 and 1190818, and Millennium Science Initiative Program  – ICN12\_009.

\end{acknowledgements}

\bibliography{myBib}

\begin{appendix}

\section{Stack excluding quiescent galaxies} \label{sec:appendix:quiescent}

\begin{figure*}
   \centering
    \includegraphics[width=0.45\linewidth]{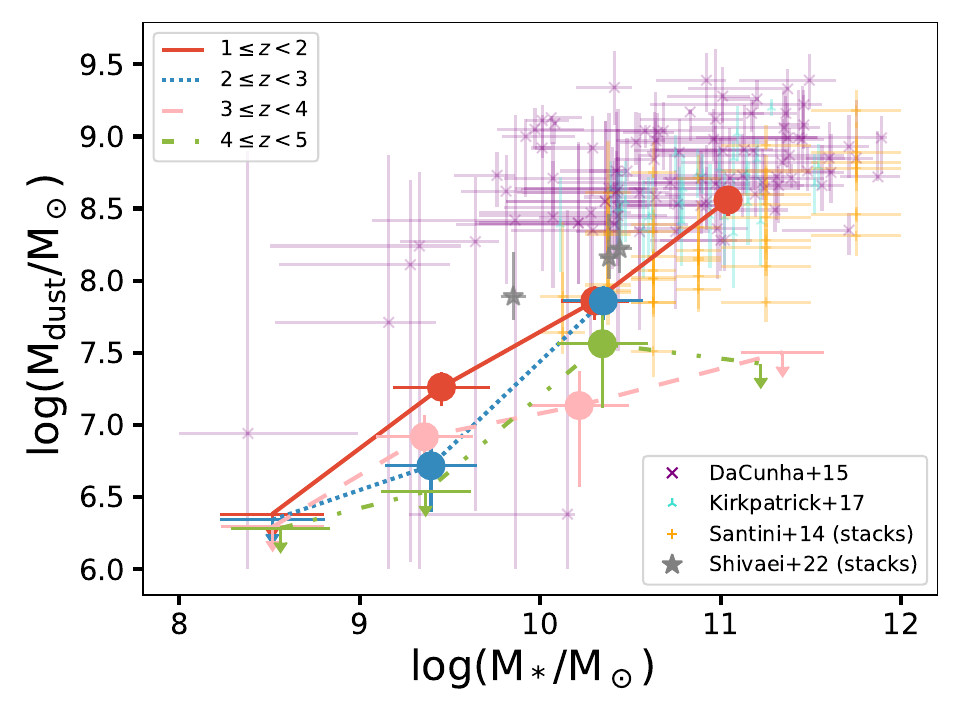}
    \includegraphics[width=0.45\linewidth]{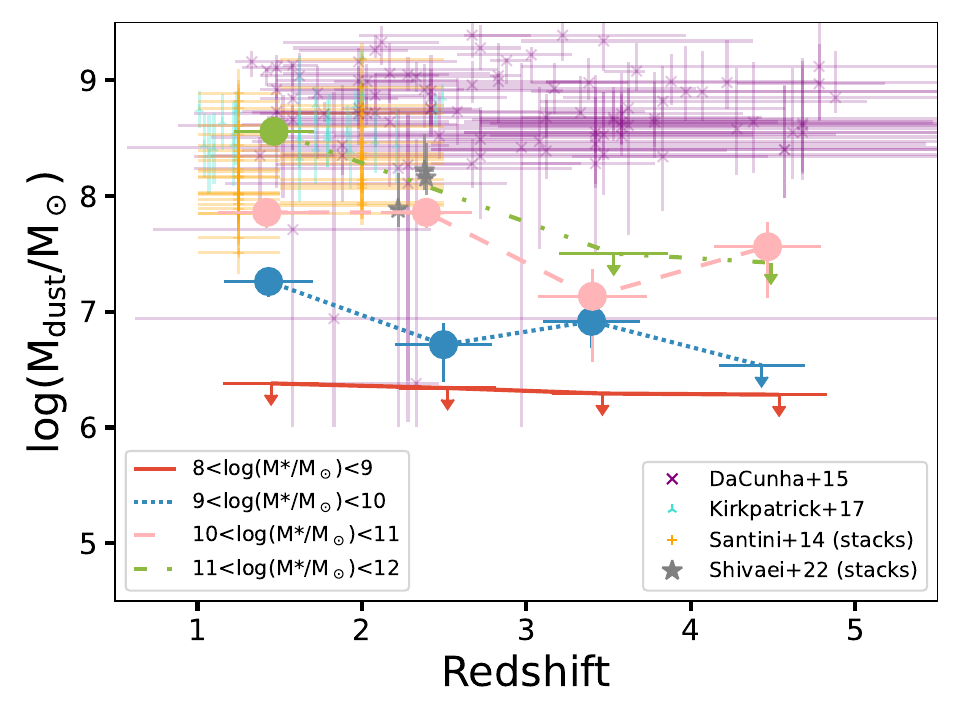}
      \caption{Similar to Figure \ref{fig:mdust_mstar} but all sources with $\Delta$MS\,$<-0.5$ are excluded. Error bars are computed in the same way as on Figure \ref{fig:mdust_mstar}. }
         \label{fig:mdust_mass_dMS}
   \end{figure*}

\begin{figure*}
   \centering
    \includegraphics[width=0.45\linewidth]{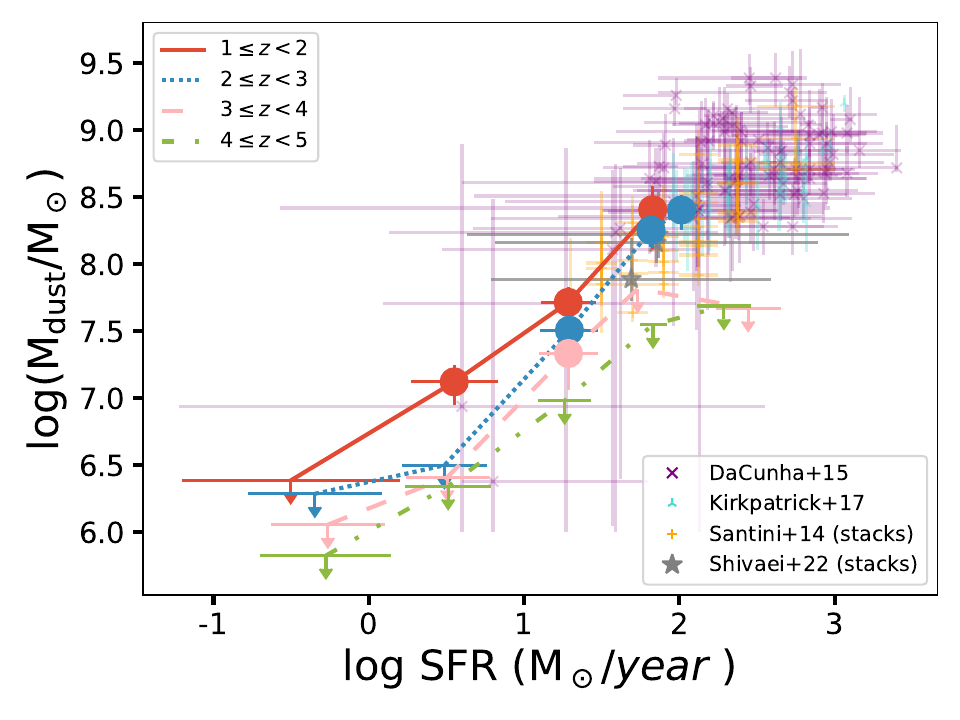}
    \includegraphics[width=0.45\linewidth]{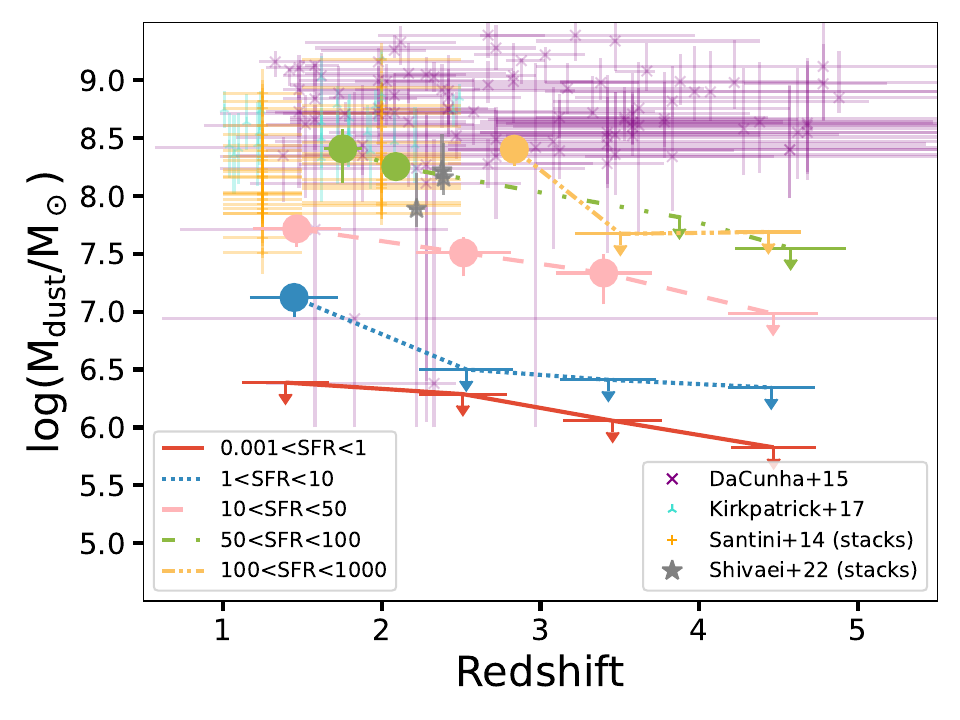}
      \caption{Similar to Figure \ref{fig:mdust_mass_dMS} but for SFR instead of stellar mass.
              }
         \label{fig:mdust_sfr_dMS}
   \end{figure*}

As shown on Figure \ref{fig:dMS} the sample is biased toward sources below the main sequence. To make sure that quiescent galaxies do not bias our analysis and the derived average dust masses, we repeat our stacking analysis after removing all sources with $\Delta$MS $<-0.5$. In total, 2157 galaxies are removed from the sample, leading to a final size of 1946 after removing quiescent galaxies. The results are shown on Figures \ref{fig:mdust_mass_dMS} and \ref{fig:mdust_sfr_dMS}.  

The main difference appear on the stellar mass subsamples, where the derived dust masses are overall higher --especially in the high-stellar mass regime-- after excluding quiescent galaxies (although some bins appear now undetected, likely due to the lower number of sources stacked). This suggest that quiescent galaxies may be overall more dust-poor at given stellar mass, driving the average down when included in the entire sample. In any event, the overall trends observed in the main analysis seem to be confirmed even when removing sources with $\Delta$MS$<-0.5$. One can note that our dust mass measurements are also in better agreement with previous individual results when quiescent galaxies are excluded from the stacks.   

\section{Fit results}

\begin{figure*}
   \centering
    \includegraphics[width=0.45\linewidth]{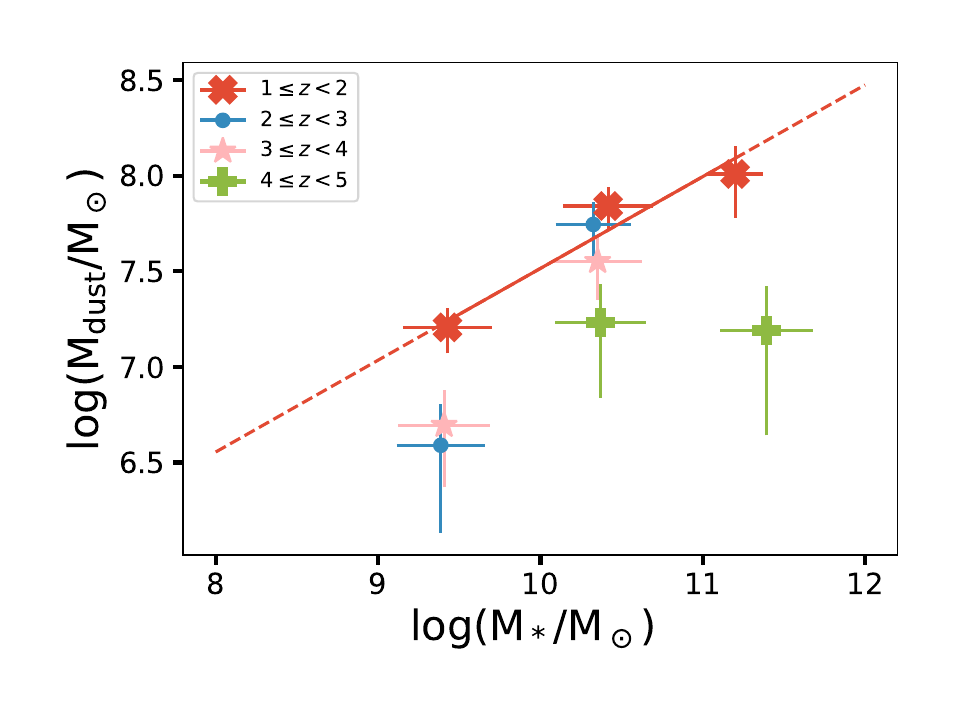}
    \includegraphics[width=0.45\linewidth]{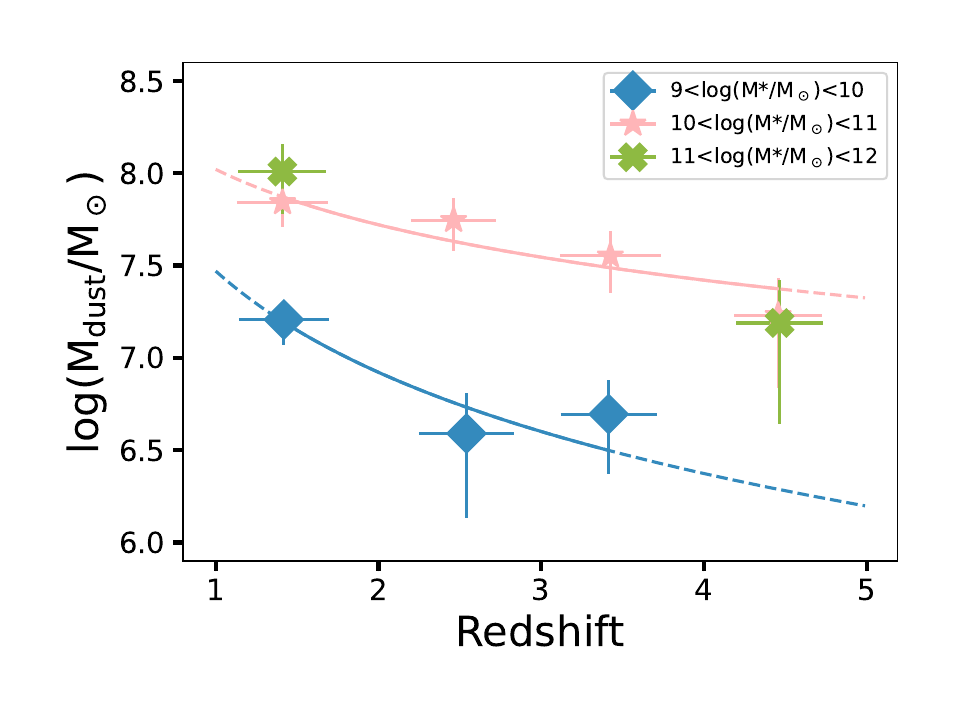}
      \caption{Similar to Figure \ref{fig:mdust_mstar} the corresponding fits are also plotted. Only detections (S/N above $3 \sigma$) are shown on the plot, and fits are only computed for subsamples with three or more data points (similarly for stellar mass subsample, right panel). Fits are plotted with a continuous (dashed) line when within (outside) the data points range.
              }
         \label{fig:mdust_mass_fit}
   \end{figure*}

\begin{figure*}
   \centering
    \includegraphics[width=0.45\linewidth]{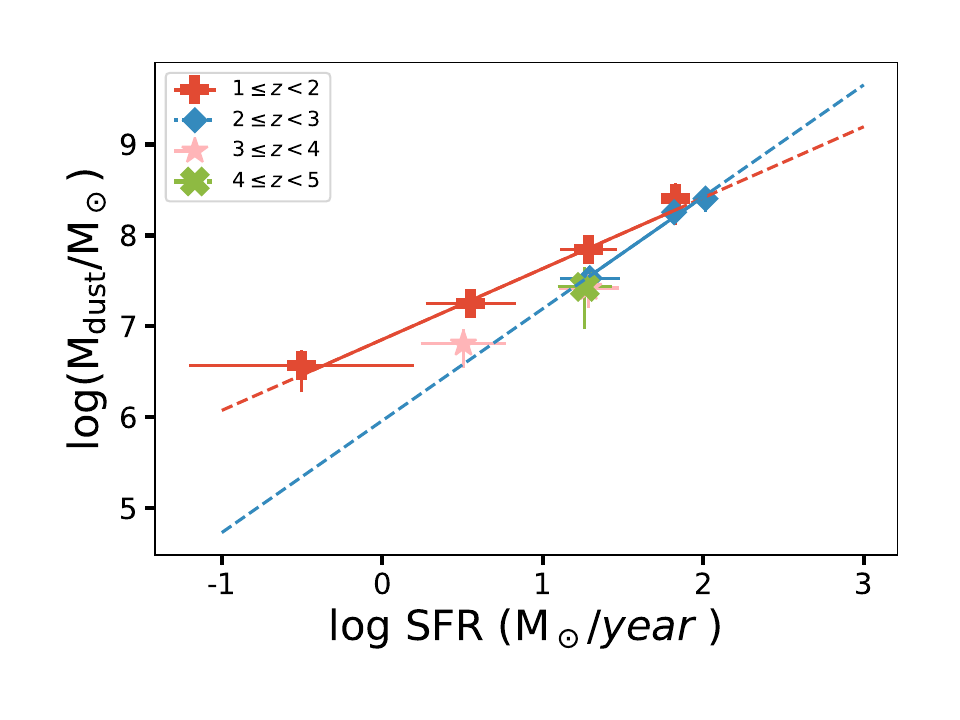}
    \includegraphics[width=0.45\linewidth]{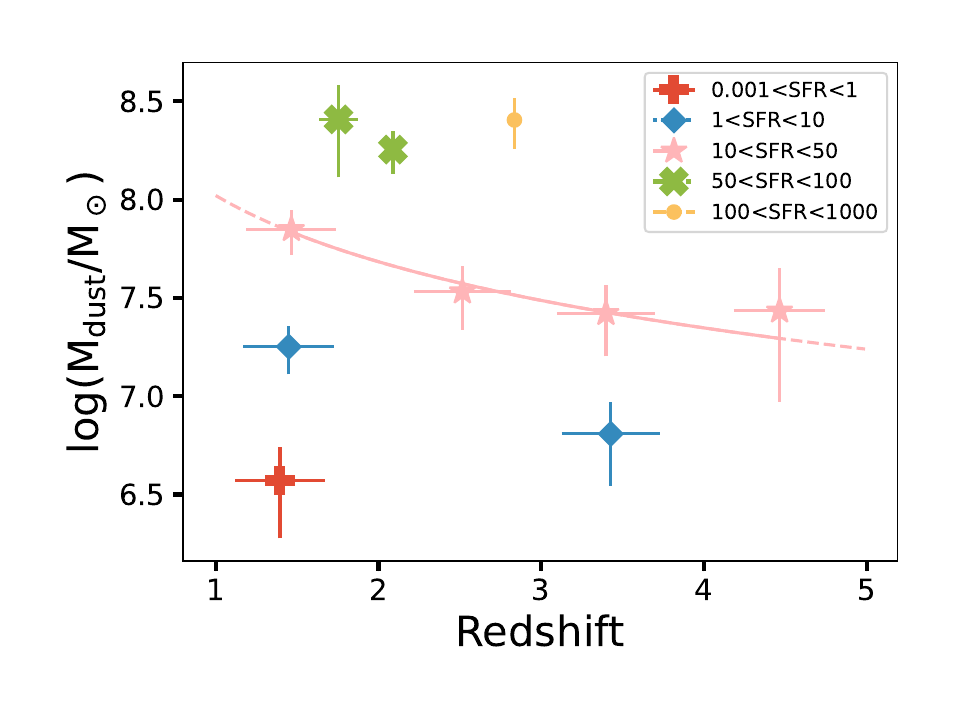}
      \caption{Similar to Figure \ref{fig:mdust_mass_fit} but for SFR instead of stellar mass.
              }
         \label{fig:mdust_sfr_fit}
   \end{figure*}

On Figures \ref{fig:mdust_mass_fit} and \ref{fig:mdust_sfr_fit} are shown the average dust mass recovered as function of stellar mass, SFR or redshift, superposed with the corresponding fit. Only the detections (i.e. S/N above $3\sigma$, see Section \ref{sec:method}) are shown and used in the fit. The coefficients of the fits are shown on Table \ref{tab:fit}.

\section{Distribution of SFR and stellar mass in the subsamples}

\begin{figure*}
   \centering
    \includegraphics[width=0.9\linewidth]{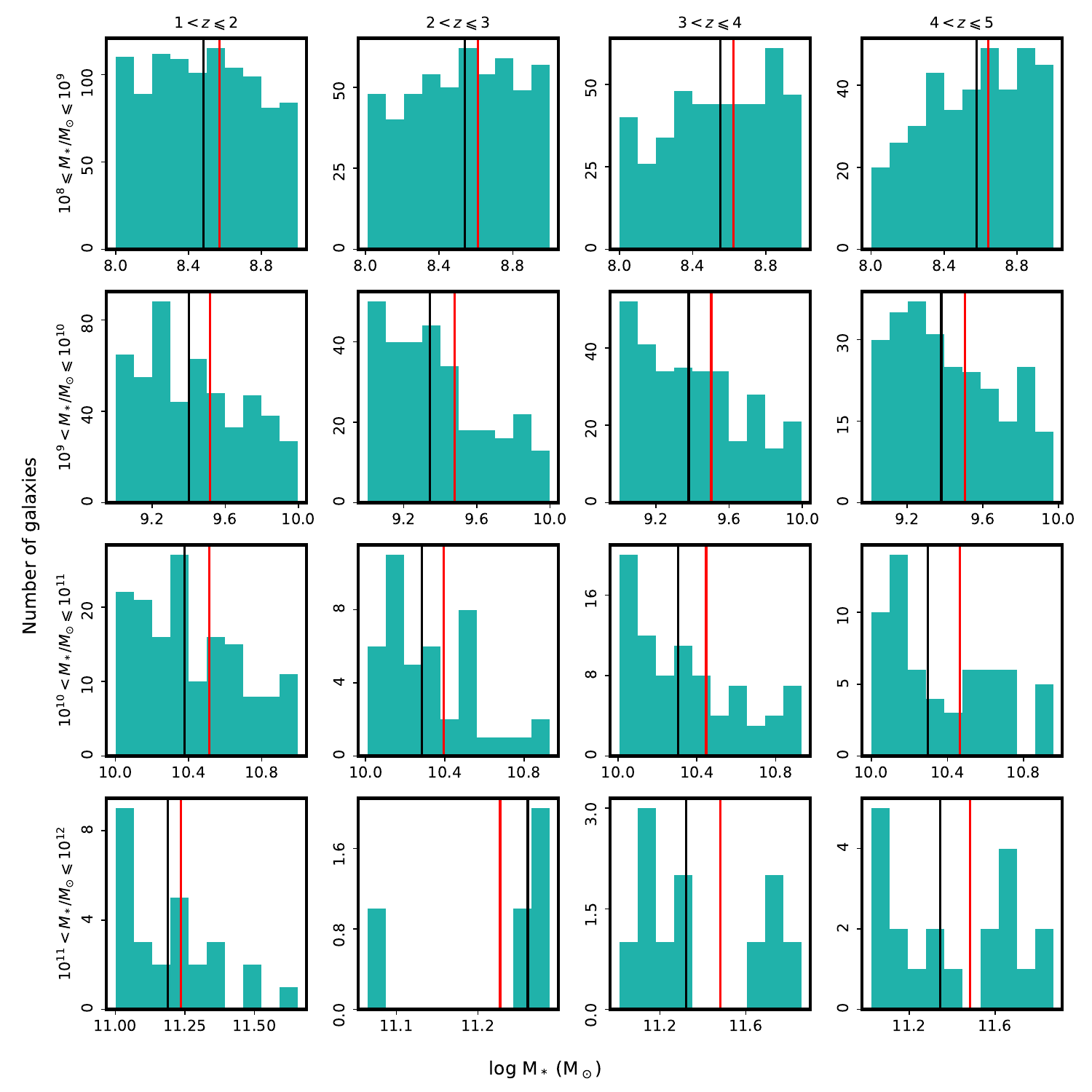}
      \caption{Distribution of the (log) stellar mass of galaxies in each subsamples. The red (black) vertical line indicates the mean (median) stellar mass of the subsample.
              }
         \label{fig:mass_subsample_hist}
   \end{figure*}

\begin{figure*}
   \centering
    \includegraphics[width=0.9\linewidth]{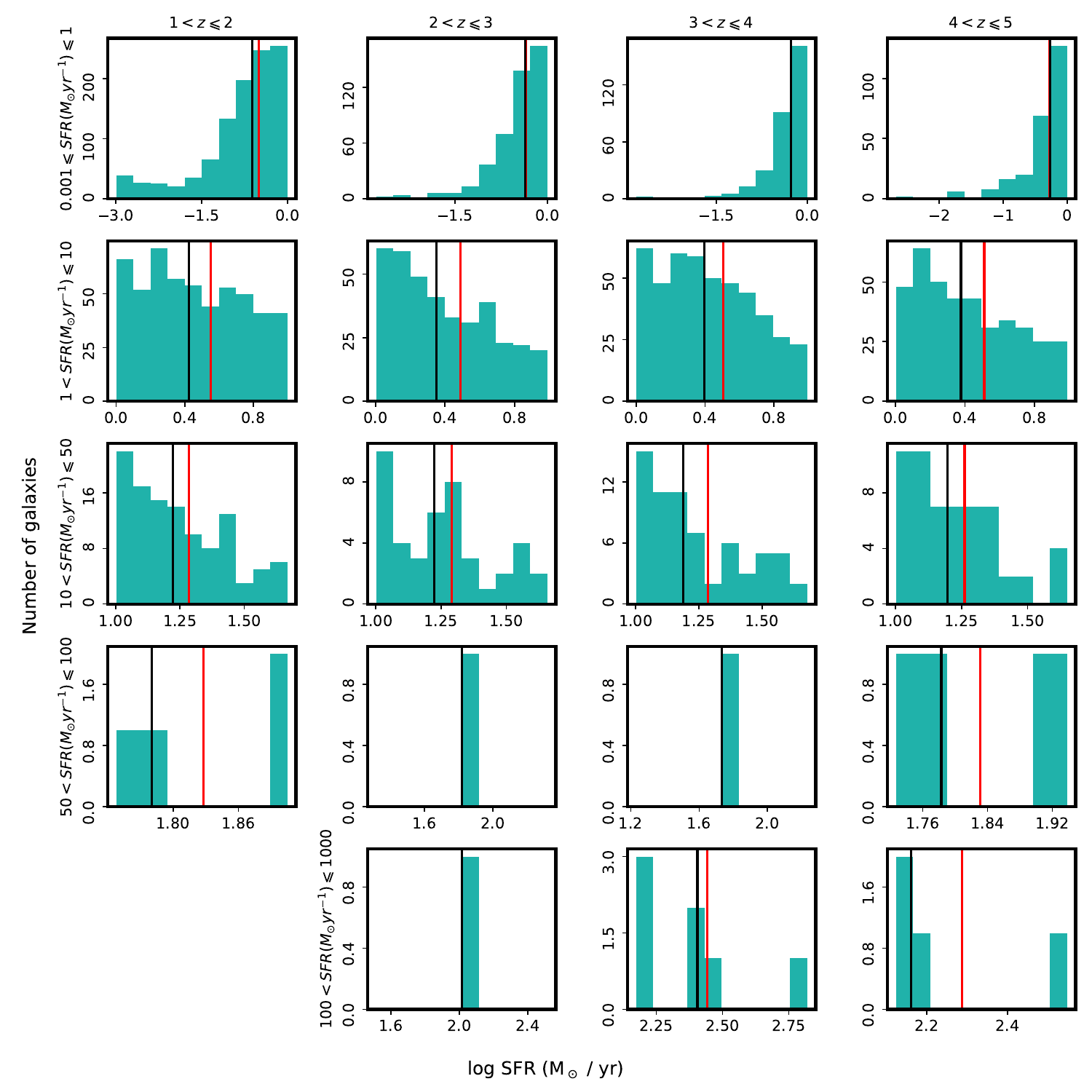}
      \caption{Similar to Figure \ref{fig:mdust_mass_fit} but for SFR instead of stellar mass. 
              }
         \label{fig:sfr_subsample_hist}
   \end{figure*}

On Figures \ref{fig:mass_subsample_hist} and \ref{fig:sfr_subsample_hist} are shown the distribution of stellar mass and SFR in the different subsamples. The mean and median values for each subsamples are also shown as red and black vertical lines respectively.

\section{Bootstrapping results}

\begin{figure*}
   \centering
    \includegraphics[width=0.9\linewidth]{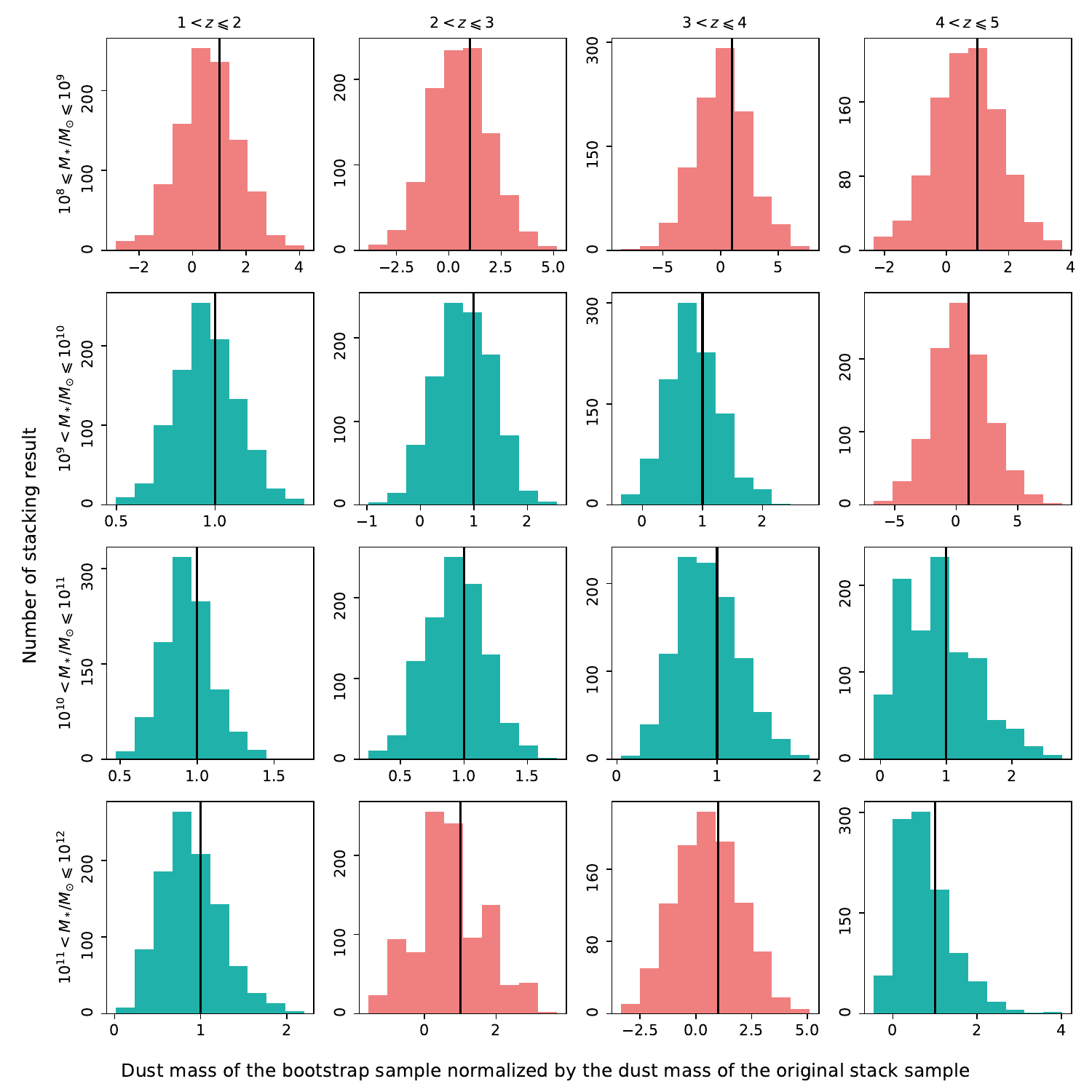}
      \caption{Distribution of the bootstrap stack results, normalized by the dust mass of the original stack. Black vertical lines at x=1 indicate the expected location of the peak distribution. Histograms in cyan (red) represents the subsample for which the signal was higher (lower) than $3 \sigma$.
              }
         \label{fig:boot_mass}
   \end{figure*}

\begin{figure*}
   \centering
    \includegraphics[width=0.9\linewidth]{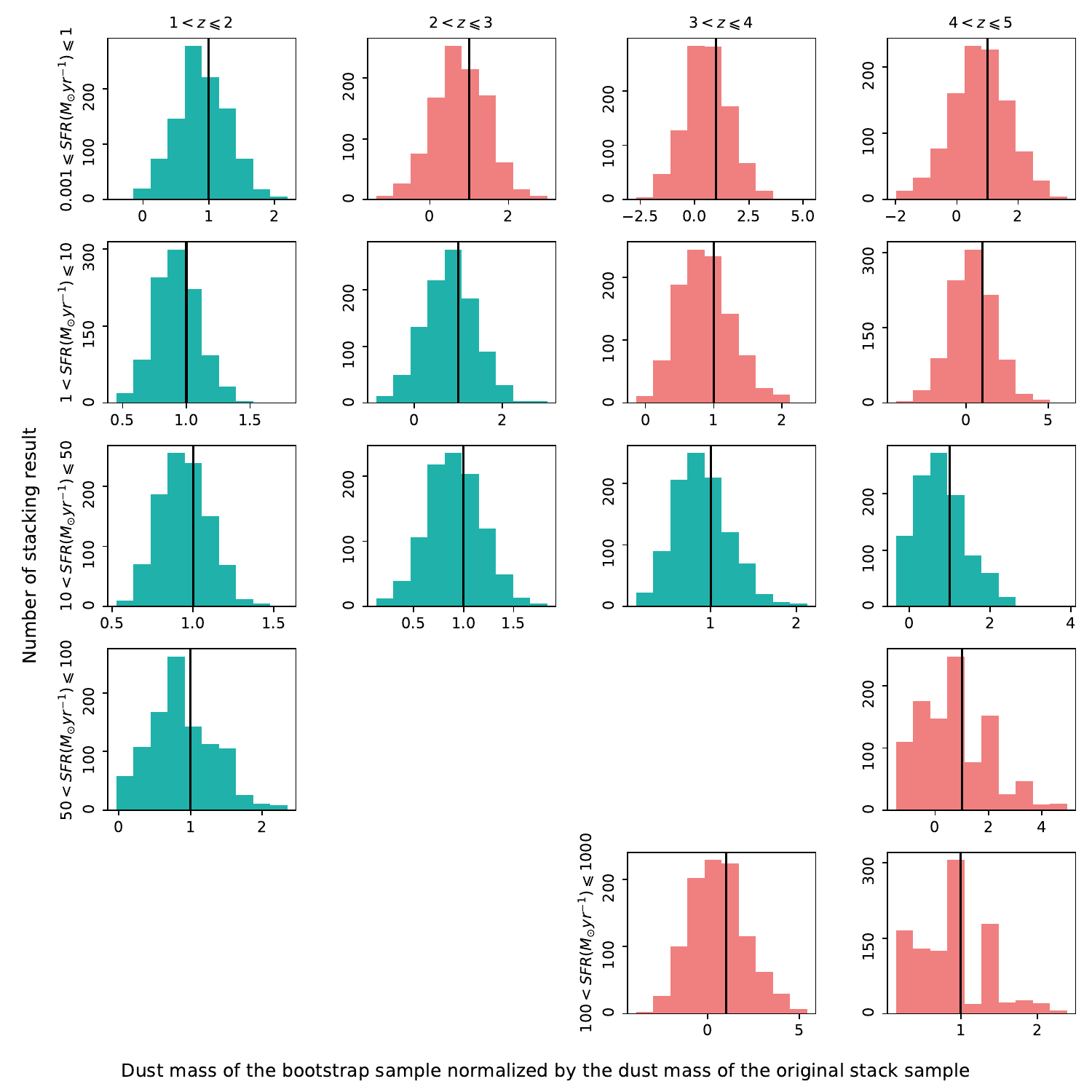}
      \caption{Similar to Figure \ref{fig:boot_mass} but for the SFR subsamples. The 3 subsamples containing only 1 source were excluded from this analysis.}
         \label{fig:boot_sfr}
   \end{figure*}

On Figures \ref{fig:boot_mass} and \ref{fig:boot_sfr} we show the distribution of the results from our bootstrapping analyses. For each subsample, 1000 new subsamples are randomly selected \citep[see][for a detailed description of the bootstrapping tool included in \textsc{LineStacker} and used here]{Jolly2020}, and their stack values are collected. The distribution of these stack results (normalized to the original stack result of the corresponding subsample) are shown. Color coding is used to distinguish between the detected (in cyan) and not detected (in red) subsamples. A vertical line is placed at $x=1$ to guide the eye.

\section{Dust mass inferred from median stacking} 

\begin{figure*}
   \centering
    \includegraphics[width=0.45\linewidth]{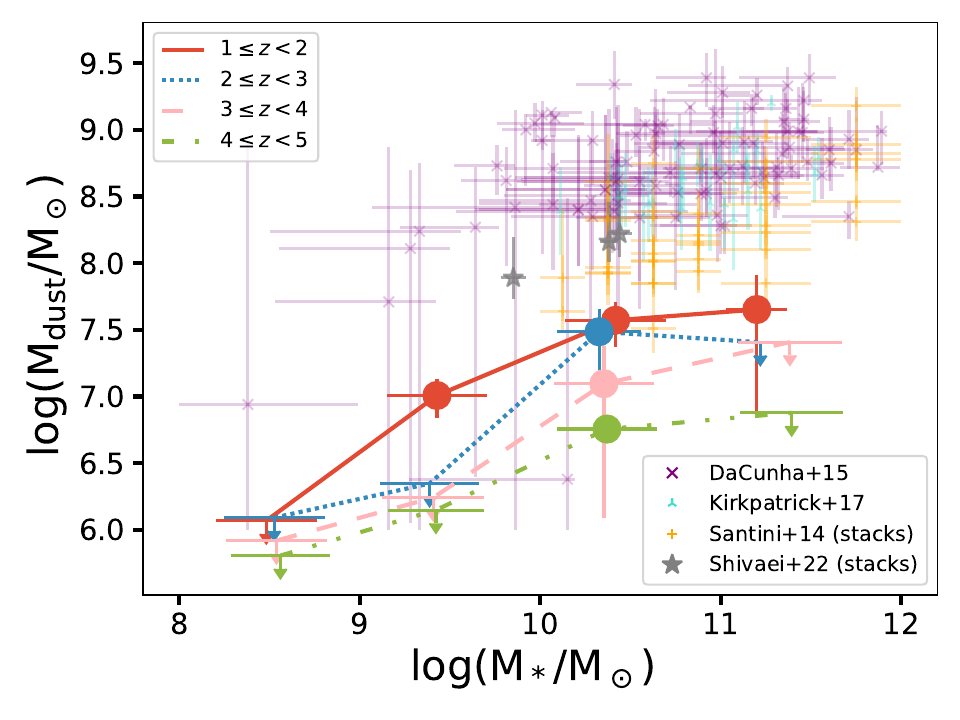}
    \includegraphics[width=0.45\linewidth]{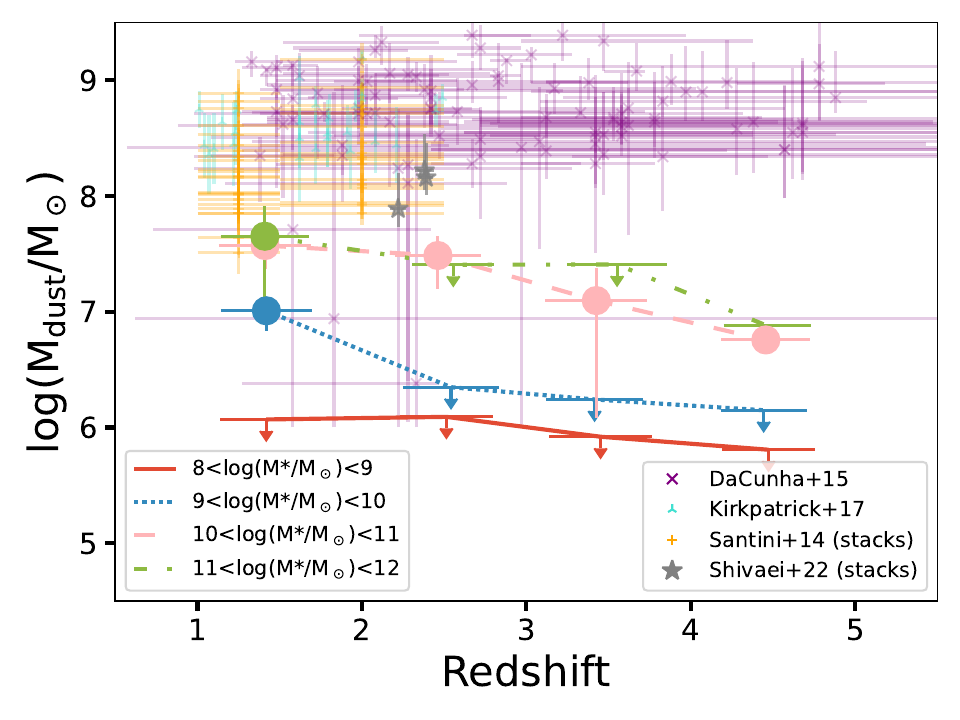}
      \caption{Similar to Figure \ref{fig:mdust_mstar} but using median stacking. }
         \label{fig:mdust_mass_median}
   \end{figure*}

\begin{figure*}
   \centering
    \includegraphics[width=0.45\linewidth]{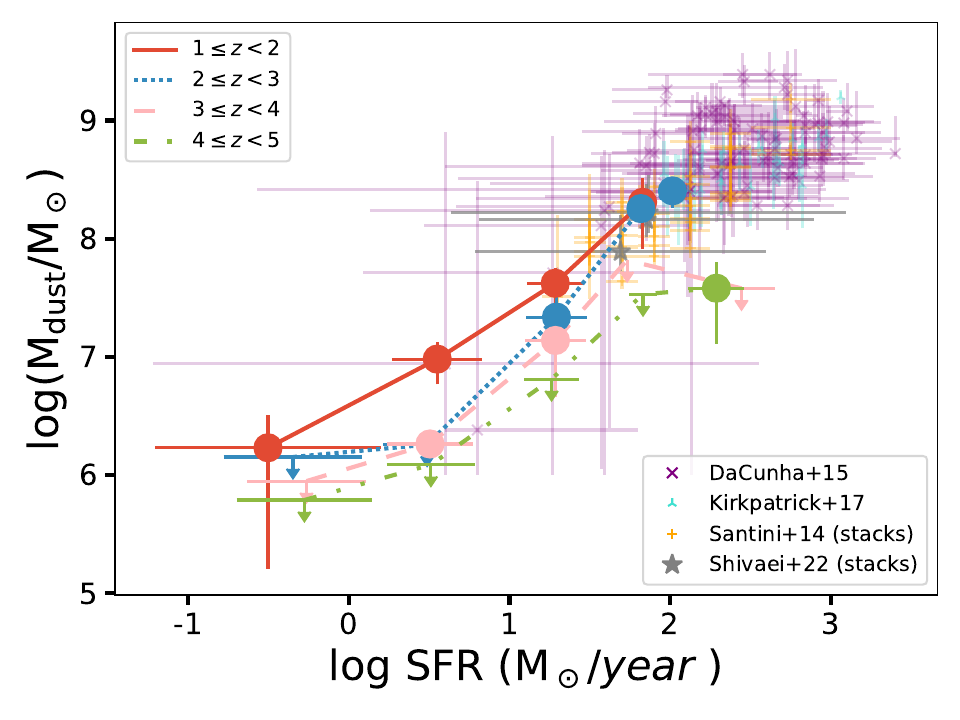}
    \includegraphics[width=0.45\linewidth]{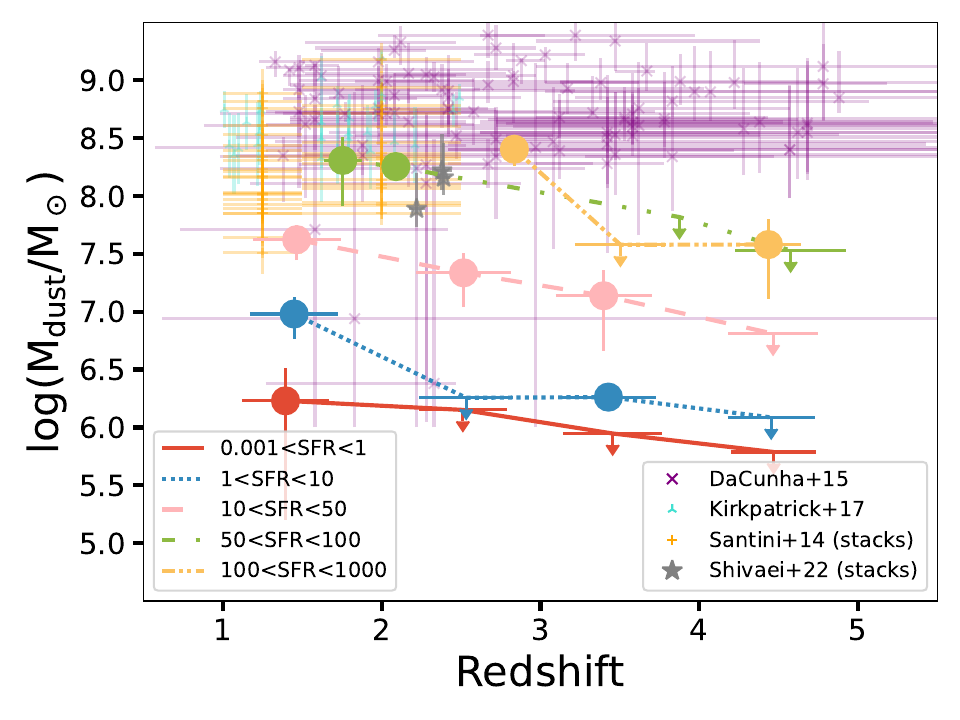}
      \caption{Similar to Figure \ref{fig:mdust_mass_median} but for SFR instead of stellar mass.
              }
         \label{fig:mdust_sfr_median}
   \end{figure*}

In addition to the mean stacking analyses we performed median stacking for each of the subsamples, in the same way as described in the main analysis. The median stacking stamps are shown on Figure \ref{fig:stamps_median} and here we show the dust mass evolution as a function of stellar mass and redshift (Figure \ref{fig:mdust_mass_median}) and as a function of SFR and redshift (Figure \ref{fig:mdust_sfr_median}), similar to Figures \ref{fig:mdust_mstar} and \ref{fig:mdust_sfr}. While the median stacking analyses show typically lower dust masses (indicating possible skewed distributions) the results and trends are overall very consistent with the mean stacking analyses.

\end{appendix}
\end{document}